\documentclass[iop,usenatbib]{emulateapj}

\usepackage{journals}
\usepackage{hyperref}
\usepackage{graphicx}
\usepackage{color}
\usepackage{epstopdf}
\usepackage{xspace}
\usepackage{threeparttable}
\usepackage{amsmath}

\definecolor{urlcolour}{rgb}{0,0,0}
\definecolor{citecolour}{rgb}{0,0,0}
\definecolor{linkcolour}{rgb}{0,0,0}
\hypersetup{colorlinks,breaklinks, urlcolor=urlcolour, linkcolor=linkcolour, citecolor=citecolour}

\newcommand{\msun}{\ensuremath{M_{\sun}}\xspace}
\newcommand{\lsun}{\ensuremath{L_{\sun}}\xspace}
\newcommand{\rsun}{\ensuremath{R_{\sun}}\xspace}
\newcommand{\ergs}{\ensuremath{{\rm erg\,s}^{-1}}\xspace}

\newcommand{\Tasc}{\ensuremath{T_{\rm asc}}\xspace}
\newcommand{\asini}{\ensuremath{a\sin i / c}\xspace}
\newcommand{\porb}{\ensuremath{P_{\rm orb}}\xspace}
\newcommand{\xmm}{{\em XMM-Newton}\xspace}

\shorttitle{A slow luminous extragalactic X-ray pulsar}
\shortauthors{Zolotukhin et al.}

\begin{document}
\title{The slowest spinning X-ray pulsar in an extragalactic globular cluster}
\author{Ivan Yu. Zolotukhin\altaffilmark{1,2,3}, Matteo Bachetti\altaffilmark{4},
Nicola Sartore\altaffilmark{5}, Igor V. Chilingarian\altaffilmark{6,2}, \and
Natalie A. Webb\altaffilmark{5,1}}
\affil{$^1$ Universit\'e de Toulouse; UPS-OMP, IRAP, 9 avenue du Colonel Roche, BP 44346, F-31028 Toulouse Cedex 4, France}
\affil{$^2$ Sternberg Astronomical Institute, Moscow State University, Universitetskij pr., 13, 119992, Moscow, Russia}
\affil{$^3$ Special Astrophysical Observatory of the Russian Academy of Sciences, Nizhnij Arkhyz 369167, Russia}
\affil{$^4$ INAF/Osservatorio astronomico di Cagliari, via della Scienza 5, I-09047 Selargius, Italy}
\affil{$^5$ CNRS, IRAP, 9 avenue du Colonel Roche, BP 44346, F-31028 Toulouse Cedex 4, France}
\affil{$^6$ Smithsonian Astrophysical Observatory, 60 Garden St. MS09, Cambridge, MA, 02138, USA}

\begin{abstract}
Neutron stars are thought to be born rapidly rotating and then exhibit a phase of a rotation-powered pulsations as they slow down to 1--10~s periods.
The significant population of millisecond pulsars observed in our Galaxy is explained by the {\em recycling} concept: during an epoch of accretion from a donor star in a binary system, the neutron star is spun up to millisecond periods.
However, only a few pulsars are observed during this recycling process, with relatively high rotational frequencies.
Here we report the detection of an X-ray pulsar with $P_{\rm spin} = 1.20$~s in the globular cluster B091D in the Andromeda galaxy, the slowest pulsar ever found in a globular cluster.
This bright (up-to 30\%\ of the Eddington luminosity) spinning-up pulsar, persistent over the 12 years of observations, must have started accreting less than 1~Myr ago and has not yet had time to accelerate to hundreds of Hz.
The neutron star in this unique wide binary with an orbital period $P_{\rm orb} = 30.5$~h in a 12~Gyr old, metal rich star cluster, accretes from a low mass, slightly evolved post-main sequence companion.
We argue that we are witnessing a binary formed at relatively recent epoch by getting a $\sim$0.8\msun star in a dynamical interaction -- a viable scenario in a massive dense globular cluster like B091D with high global and specific stellar encounter rates.
This intensively accreting non-recycled X-ray pulsar provides therefore a long-sought missing piece in the standard pulsar recycling picture.
\end{abstract}


\section{Introduction}
\label{sec_intro}

Around 2000 pulsars are known, where the majority of these are 'regular' pulsars which have pulse periods between tens of milliseconds to approximately a second, and magnetic field strengths of $\sim$10$^{12}$ G \citep{manchester05}. These pulsars show a general spin down to longer periods. However, a few hundred of these pulsars show much shorter periods, of the order a millisecond, along with lower magnetic fields of $\sim$10$^8$ G. It is believed that these millisecond pulsars (MSPs) are the descendants of neutron stars/pulsars found in X-ray binaries.  Accretion onto a neutron star from
a close companion is believed to transfer angular momentum to the
neutron star, spinning it up to  periods of milliseconds \citep{alpa82,radh82}. This is strongly
supported by both the discovery of a MSP in an X-ray binary system
\citep[SAX 1808.4-3658,][]{wijn98} and the presence of
kilo-Hertz Quasi-Periodic Oscillations in many LMXBs, which have been
found to show millisecond pulsation periods \citep[see][and references therein]{vand98}. More recently, \cite{arch09} showed that the previously accreting millisecond pulsar FIRST J102347.67+003841.2 had ceased to accrete and radio pulsations could subsequently be observed, thus supporting the 'recycled' pulsar idea (see \citet{patruno12a} for a review).

Many of the known MSPs are found in Galactic globular clusters (GCs), where as of the end of 2015, almost 140 MSPs have been detected\footnote{\url{http://www.naic.edu/~pfreire/GCpsr.html}}.  Globular clusters are dense spherical systems of $\sim$10$^4$--10$^6$ old stars \citep[e.g.][]{heno61}.  Their old age implies that they should also contain many compact objects \citep[e.g.][]{hut92}. Stellar encounters, which are extremely rare in lower density
regions, can occur in globular clusters on time-scales comparable with
or less than the age of the Universe.  This would indicate that many
Galactic globular cluster stars have undergone at least one encounter
in its lifetime.  Encounters between stars is one way in which
binaries can be produced.  The encounter rate ($\Gamma$) due to tidal capture
\citep{fabi75} is proportional to the encounter cross-section, the
relative velocity of the stars and the number density of stars
in the cluster (core).  Both primordial binary systems and those formed due to encounters
should exist in globular clusters due to the dense environments, but encounters between a binary and either a single star or a
binary system would more readily occur as the cross sections
are significantly larger, thus increasing the likelihood of an
encounter.  This explains the large number of recycled pulsars that we observe in GCs. 

However, a small number (6) have periods greater than 0.1~s and these pulsars have not yet been fully recycled. The longest of these, B 1718-19 in NGC~6342, has a period of 1.004~s and a magnetic field of $\sim10^{12}$ G, typical of a 'regular' pulsar.  It appears to be a young pulsar, with a characteristic age of 1 $\times$ 10$^7$ years \citep{lyne93}.  \cite{lyne93} propose that this young pulsar originated either from a collision between an old neutron star and a cluster star, or that a white dwarf accreting from a companion underwent an accretion induced collapse \citep{mich87}. 
\citet{verbunt14} suggest that the current main-sequence companion of an old neutron star in that system has replaced the original one in an exchange encounter.

Following the recent detection of coherent pulsations from an ultra-luminous X-ray source (ULX) in the M82 galaxy ($\simeq$3.5~Mpc away) using {\it NuSTAR} data \citep{bachetti14}, which showed that this bright source was in fact a neutron star, we started to search archive {\em XMM-Newton} data to find similar sources, in order to address questions such as how can such super-Eddington luminosities be possible in a neutron star \citep[see e.g.][for further discussion]{Mushtukov+15,KingLasota16}. 

In this paper we describe the analysis that made it possible to detect a 1.2~s pulsar in the X-ray binary 3XMM J004301.4+413017,  associated with the globular cluster B091D from the Revised Bologna Catalog of \object{M31} globular clusters \citep[RBC V.5;][]{galleti04}, using public data. For brevity we denote the pulsar XB091D after its host globular cluster designation. This is the first persistently accreting X-ray pulsar hosting a neutron star detected in \object{M31}\footnote{Other known X-ray pulsating sources in \object{M31} are a transient pulsar candidate XMMU J004415.8+413057 in a high-mass X-ray binary with a period of 197~s \citep{trudolyubov05}, and two supersoft sources powered by accreting white dwarfs: XMMU J004252.5+411540 with a period of 217.7~s \citep{trudolyubov08}, and XMMU J004319.4+411759 with a period of 865.5~s \citep{osborne01}.}, and it also has the longest period among all known pulsars (rotation-powered and accreting) in globular clusters, being more than an order of magnitude slower than the mildly recycled accreting pulsar from Terzan~5 globular cluster \citep{papitto11}.  We note that this pulsar has also recently been detected by \cite{esposito16}, but these authors interpret the nature of this source quite differently. In this paper, we discuss possible evolutionary scenarios that may produce such slowly rotating neutron stars in globular clusters. 

The content is organized as follows: Section \ref{sec:data} briefly describes the \xmm photon database (which will be described in detail elsewhere) and the dataset used for the initial pulsation detection, as well as the \xmm pulsar factory analysis methods which resulted in the automated detection of the pulsed X-ray emission; Section \ref{sec:bsearch} covers the manual blind search for pulsed emission in all available \xmm data for this source and the determination of source's orbital parameters; Section \ref{sec:timing} and \ref{sec:spec} summarize our findings on the timing and spectral properties of this source which are then discussed in Section \ref{sec:discussion} where we also argue on the possible origin and evolution of this system.

\section{Pulsations search data and methods}
\label{sec:data}

\subsection{Photon database}

\begin{figure*}
\plotone{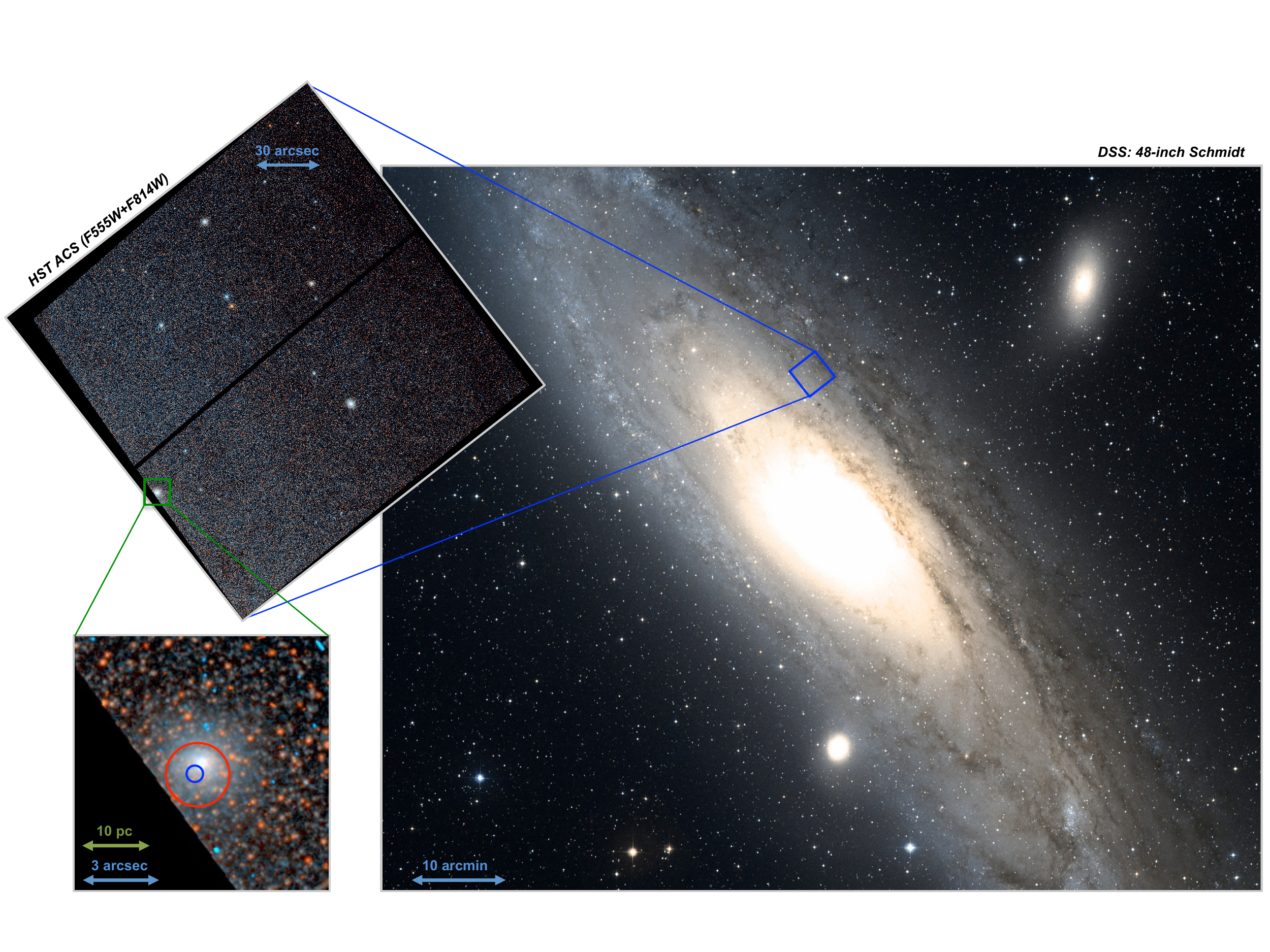}
\caption{The position of the XB091D and its host globular cluster B091D in the Andromeda galaxy in several scales -- using colored images from the Digitized Sky Survey and {\it HST} ACS.
On the zoomed inset in the bottom left corner, we overplot the 95\% confidence level X-ray positional uncertainties from the {\it Chandra} (red circle) and {\it XMM-Newton} (blue circle) catalogs.}
\label{fig:m31}
\end{figure*}

For this study we created a database of all photons registered by the EPIC/pn detector of the \xmm satellite and operated in imaging mode during the 7781 observations that took place between 2000 and mid 2013. These are the same observations that were used to compile the 3XMM-DR5 catalog \citep{rosen15}.
The photon data were taken from the event lists called {\tt PIEVLI} files publicly available from the \xmm science archive\footnote{\url{http://xmm.esac.esa.int/xsa/}}.
These are science-ready data products that come from the pipeline run by the \xmm Survey Science Centre (XMM SSC)\footnote{\url{http://xmmssc.irap.omp.eu}}. 
Each file is a binary table with one row per event that contains the following information: (a) event time (satellite clock), (b) raw CCD pixel and projected sky coordinates relative to the nominal pointing position, (c) corrected and uncorrected event energy; (d) event quality flag; (e) CCD number.
Each {\tt PIEVLI} file also contains good time intervals determined separately for each pn CCD by the XMM SSC pipeline software.
We point out that it is not necessary to retrieve the complete \xmm observation archive (ODF archive) for the massive scale timing analysis with the \xmm.

These event lists represent the lowest and most detailed data level we use for initial timing studies.
These are accompanied by higher level data in the form of the 3XMM-DR5 catalog of X-ray sources and their individual observations acting essentially as an index for the navigation within the photon database.
The connection between these two very different sets of information is achieved by means of the known transformation between pixel coordinates of the events and world coordinates of the X-ray sources from the 3XMM-DR5 catalog.
We used the World Coordinate System (WCS) transformation data available for every EPIC/pn exposure in the {\tt PIEVLI} file.
In this fashion one can make event extraction for any detection from the 3XMM-DR5 catalog using its world coordinates, right ascension and declination.

The catalog of X-ray sources was cross-matched with other catalogs of astrophysical objects in order to find possible counterparts and determine the source type or location within a galaxy.
This gives us the possibility to easily extract arbitrary photon lists of e.g. all photons coming from the \object{M31} galaxy, or all photons from known magnetars.

The last stage before being able to launch our timing analysis codes over the sequence of extracted photon lists is the barycentric correction, i.e. correcting the event times to the Solar System barycenter.
For this purpose we used the \xmm orbit files fed to the {\it barycen} task from the \xmm\ {\sc SAS} version 13.5 software and the source positions from the 3XMM-DR5 catalog.

We note that it is now possible to access this photon database as well as the 3XMM-DR5 catalog data through the convenient web interface\footnote{\url{http://xmm-catalog.irap.omp.eu}} that we developed while working on the 3XMM-DR5 catalog compilation and this project.
In particular, one can extract barycentered photons in arbitrary regions from the observation level event lists using nothing but a web browser.
More details on this website for the quick-look science analysis of the \xmm data will be presented in a separate paper (Zolotukhin et al., in prep.)

\subsection{Pulsar factory analysis software}

The photon database provides a way to easily extract calibrated and barycentered event lists for arbitrary sets of astrophysical sources observed with the EPIC/pn camera on board the \xmm satellite.
We developed an analysis software aimed at finding coherent pulsations in the photon database, optimized for high throughput.

It consists of few analysis layers.
First, for each event list it produces the power density spectrum (PDS) adapting the time binning so that the Nyquist frequency is well above 2 kHz and the total number of bins between gaps in the light curve is factorable with small prime numbers for better performance with the {\tt numpy} FFT algorithm. 
In the PDS the algorithm searches for peaks exceeding the 99~per~cent detection level, including the number of trials, following the standard rules from \citet{Leahy+83}.
If such peaks are found, it launches the $Z^2_n$ test \citep{buccheri83} in the vicinity of their frequencies.
If a peak from the PDS is confirmed with the $Z^2_n$ test, an accelerated search with the PRESTO code \citep{RansomThesis} is launched to constrain further or discard this pulsation frequency candidate.
As the final step of the automated procedure the pulsar factory checks this frequency in other observations of this source from the 3XMM-DR5 catalog. If the detection is highly significant ($>$$n\sigma$, where $n$ can vary between runs) or the same frequency is found in other detections, an operator is notified.

We first tested the \xmm pulsar factory code using \xmm data on a set of known X-ray bright magnetars from the McGill Online Magnetar Catalog \footnote{\url{http://www.physics.mcgill.ca/~pulsar/magnetar/main.html}} \citep{olausen14}.
After achieving the stable detection of known coherent pulsation periods in this automated regime, we launched a larger period search on the unstudied sample of X-ray sources in the catalog. 
The \object{M31} galaxy was an obvious choice of survey region for its large number of known X-ray sources and its relatively low distance, and was included in our first survey.

\begin{deluxetable}{lc}
\tablecolumns{2}
\tablewidth{3in}
\tablecaption{Best orbital solution found in this paper. }
\tablehead{   
  \colhead{Parameter} &
  \colhead{Value}
}
\startdata
\porb		&		1.27101304(16)\,d\\
\Tasc      &       MJD 56104.791(26)\\
\asini      &       2.89(13)\,l-sec \\
$e$			&		$<$0.003
\enddata
\tablecomments{Brackets indicate 1-$\sigma$ uncertainties, as returned by TEMPO2. The upper limit on the eccentricity is the maximum uncertainty returned by the {\tt ELL1} model, for values of the eccentricity always consistent with 0.\label{tab:orb}}
\end{deluxetable}

\begin{deluxetable*}{lcccccc}  
\tablecolumns{7}
\tablewidth{0pt}
\tablecaption{\xmm observations used for timing analysis in this study and its main results.\label{tab:obs}}
\tablehead{   
	\colhead{ObsID} & 
	\colhead{Obs. start} & 
	\colhead{Exposure} & 
	\colhead{$P_{\rm spin}$} & 
	\colhead{$S$ / d.o.f.} & 
	\colhead{Pulse depth} \\
	\colhead{} & 
	\colhead{UTC} & 
	\colhead{(s)} & 
	\colhead{(s)} & 
	\colhead{} & 
	\colhead{}
}
\startdata
0112570101 & 2002-01-06 18:45:45 & 64317 & 1.203898(49) & 8.7 & $0.33 \pm 0.07$\\
\hline \\
0405320701 & 2006-12-31 14:24:50 & 15918 & 1.203731(11) & 4.1 & $0.28 \pm 0.08$\\
0405320901 & 2007-02-05 03:44:24 & 16914 &  &  & \\
\hline \\
0505720201 & 2007-12-29 13:42:13 & 27541 & 1.203738(10) & 7.3 & $0.20 \pm 0.05$\\
0505720301 & 2008-01-08 07:01:05 & 27219 &  &  &  \\
0505720401 & 2008-01-18 15:11:47 & 22817 &  &  &  \\
0505720501 & 2008-01-27 22:28:21 & 21818 &  &  &  \\
\hline \\
0551690201 & 2008-12-30 03:27:52 & 21916 & 1.203662(8) & 5.0 & $0.23 \pm 0.05$\\
0551690301 & 2009-01-09 06:19:54 & 21918 &  &  &  \\
0551690501 & 2009-01-27 07:23:03 & 21912 &  &  &  \\
0551690601 & 2009-02-04 13:21:03 & 26917 &  &  &  \\
\hline \\
0600660201 & 2009-12-28 12:42:54 & 18820 & 1.203675(13) & 4.2 & $0.25 \pm 0.08$\\
0600660501 & 2010-01-25 02:39:14 & 19715 &  &  &  \\
\hline \\
0650560301 & 2011-01-04 18:10:16 & 33415 & 1.203651(8) & 9.1 & $0.31 \pm 0.06$\\
0650560401 & 2011-01-15 00:16:57 & 24316 &  &  &  \\
0650560601 & 2011-02-03 23:58:12 & 23918 &  &  &  \\
\hline \\
0674210201 & 2011-12-28 01:07:36 & 19034 & 1.203634(13) & 4.9 & $0.35 \pm 0.09$\\
0674210301 & 2012-01-07 02:47:01 & 15433 &  &  &  \\
0674210401 & 2012-01-15 15:00:38 & 19916 &  &  &  \\
0674210501 & 2012-01-21 12:22:03 & 17317 &  &  &  \\
\hline \\
0690600401 & 2012-06-26 06:29:43 & 122355 & 1.203698(4) & 20.6 & $0.23 \pm 0.03$\\
0700380501 & 2012-07-28 15:16:27 & 11914 &  &  &  \\
0700380601 & 2012-08-08 23:08:08 & 23916 &  &  &  \\
\hline \\
0701981201 & 2013-02-08 22:19:55 & 23918 & 1.20373(13) & 7.0 & $0.45 \pm 0.14$
\enddata
\tablecomments{
Available observations were split into 9 blocks which were analysed with the assumption that NS spin period $P_{\rm spin}$ does not change much within them.
$S$ / d.o.f. is the statistical significance of the obtained solution as defined in the Appendix~\ref{sec:epoch_folding}.
Pulse depth (also referred as pulsed fraction) is defined as quantity $A$ there as well.
Uncertainties represent $1\sigma$ confidence interval.
}
\end{deluxetable*}

\section{Detection and orbital parameters}\label{sec:det}\label{sec:bsearch}

During the \object{M31} test run, the automated \xmm pulsar factory algorithm detected pulsations, at about the same period, in 3 observations (ObsIDs: 0112570101, 0505720301, P0650560301) of the source 3XMM J004301.4+413017\footnote{See the source web page at the 3XMM-DR5 catalog website: \url{http://xmm-catalog.irap.omp.eu/source/201125706010086}}.
The $Z^2_n$ test triggered by the detection confirmed the candidate.
These detections are fully reproducible online from the source's event extraction pages, e.g. for observation 0112570101: \url{http://xmm-catalog.irap.omp.eu/pievli/101125701010068}.

In order to look for the detected pulsation in more ObsIDs, we ran an accelerated search with PRESTO in all ObsIDs containing the source. 
The very high values of period derivative required by the accelerated search, the clear improvement of detection significance when adding a second derivative in the search, and the shape of the track in the phaseogram shown by PRESTO, pointed towards the presence of orbital modulation. 
Largely following the same procedure described in \citet{bachetti14}, we cut the two longest ObsIDs into chunks, 10--30\,ks long, and ran an accelerated search with custom-made software, searching the solution in the $\nu-\dot{\nu}$ plane (where $\nu$ indicates the pulse frequency) that yielded the highest $Z^2_2$ statistics. 

Inside the long ObsIDs (0112570101, 0690600401) the best-solution frequency and frequency derivative clearly followed a sinusoidal law with a period between 1 and 2 days, as expected from orbital modulation. 
We fitted simultaneously the two values of frequency and frequency derivative in ObsID 0690600401 with the expected variation due to orbital motion, and obtained a first estimate of the orbital parameters ($\porb\sim1.2$\,d, $\asini\sim2.60$\,l-sec, $\Tasc\approx{\rm MJD} 56104.789$, where \asini is the projected semi-major axis and \Tasc the time of passage through the ascending node).
Starting from the first rough estimate of these parameters and, by trial-and-error, trying to align the pulses in the phaseogram in the chunks first, then calculating TOAs with a custom implementation of the {\tt fftfit} method \citep{Taylor92} and using TEMPO2 to fit an orbital solution with the {\tt ELL1} model\footnote{This model is appropriate for quasi-circular orbits, see \url{http://www.atnf.csiro.au/people/pulsar/tempo/ref_man_sections/binary.txt}}, we reached a solution valid to ObsID 0690600401.
We then applied the solution to the other long ObsID, refining the orbital parameters so that every residual orbital modulation was eliminated by assuming a constant spin through the observation. The solution found in this way is the following: $\porb\approx1.2695$\,d, $\asini\approx2.886$, $\Tasc\approx56104.7907$. 
Then, we addressed the ObsIDs in between, using ``quantized'' values of the orbital period that conserved the ascending node passages close to the two long ObsIDs. For every value, we calculated the scatter that it produced on the TOAs and selected the one that produced the lowest scatter. 
The eccentricity fitted by {\tt ELL1} was always consistent with 0 at the $\sim2\sigma$ level, with an uncertainty of 0.002--0.003. We use this last number as an upper limit on the eccentricity. 
The full solution is in Table \ref{tab:orb}.

We used this solution to look for pulsations in the remaining ObsIDs and refine the estimate. 
In the next section we describe this procedure.

\section{Refined timing analysis}
\label{sec:timing}

Except a few cases (ObsIDs: 0112570101, 0650560301, 0690600401), all other
individual observations yield poor photon statistics in order
to determine the pulsation period and the pulse shape with enough
statistical significance for detailed interpretation.

Therefore, we attempted the search of coherent pulsations by combining
several datasets spanning 2 to 5 months in different years of observations
by assuming that the period did not change among individual observations
within each block.  First, we corrected all the photon arrival times using
the orbital elements of the binary system reported above.  Then we used the
period grid search using the $S$ statistics with regularization (see
Appendix for details) leaving the orbital phase as an additional free
parameter.

Then, for each year we started with the first observation typically taken in
late December or early January and then started adding 
observations checking that the $S$ statistics around the probable period,
according to the increase in the exposure time suggesting that the
pulsations are still coherent.  If the $S$ statistics \citep{Leahy+83} did not increase or
decrease when we added the additional observation, we concluded that the
period changed significantly and started a new block.  The results of our
timing analysis are provided in Table~\ref{tab:obs} and the blocks of
observations used for coherent searches are separated with horizontal lines. 
We estimated uncertainties of the period measurements analytically from the
photon statistics, exposure time, and the pulse properties as explained in
the Appendix. In Fig.~\ref{fig:pulseall} we provide the 9 recovered pulse
profiles between 2002 and 2012.

We note that when correcting photon arrival times using the orbital solution of \citet{esposito16}, we were only able to get sufficient pulsations significance in three observations (0112570101, 0650560301, 0690600401), whereas in all remaining datasets the pulsating signal was not detected because it was smeared.
At the same time the orbital solution obtained in this study allows to significantly detect pulsations in all observations listed in Table~\ref{tab:obs}.
Though these two orbital parameter estimates agree on the order of magnitude with each other, the solution presented here is more precise.

\begin{figure}
\includegraphics[width=\hsize]{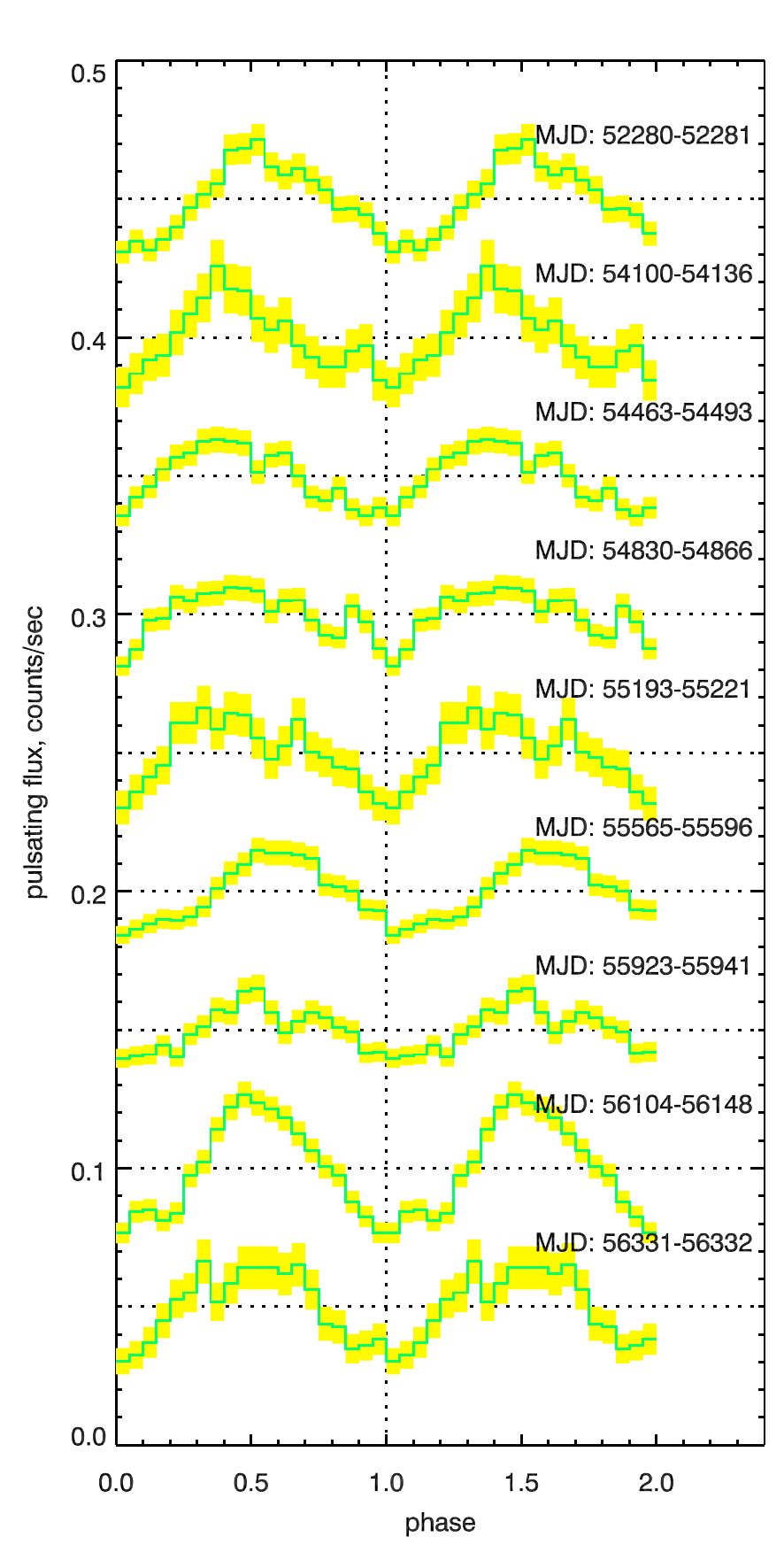}
\caption{Recovered pulse profiles of XB091D obtained from the search for
coherent pulsations using the regularized $S$ statistics in 9 combined datasets listed in Table~\ref{tab:obs}.
The shift along the Y axis is arbitrary and is made for clarity.
\label{fig:pulseall}}
\end{figure}

\begin{figure}
\includegraphics[width=\hsize]{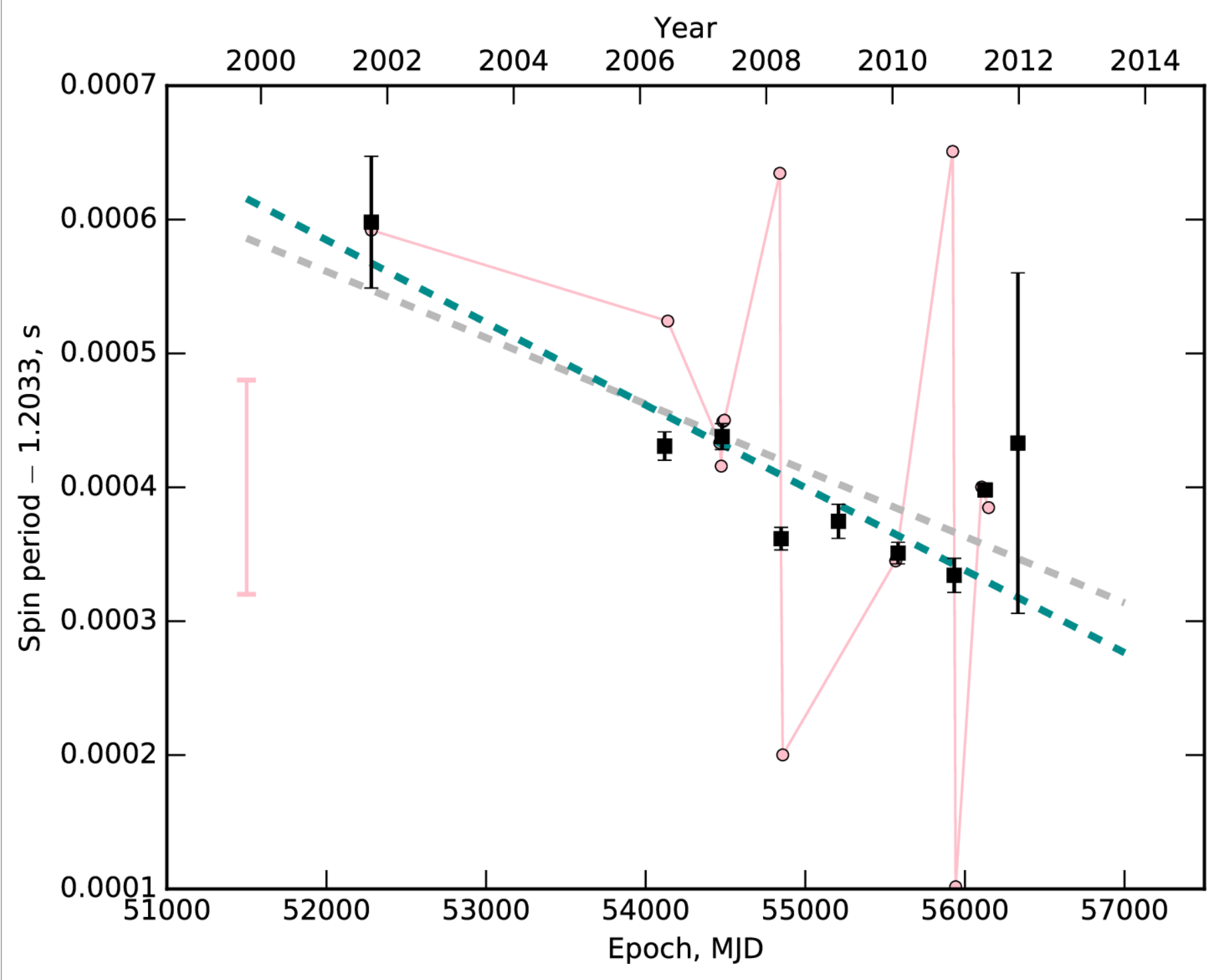}
\caption{Evolution of neutron star spin period.
Spin period estimates obtained from individual observations are displayed in light color.
They fluctuate strongly, however a typical uncertainty of individual measurement (plotted separately on the left) is large due to the poor photon statistics. What might be interpreted as a frequent change of $\dot{P}$ is, in reality, driven by these uncertainties. 
For this reason, we use the combined datasets listed in Table~\ref{tab:obs} to estimate better local period solutions. These are indicated by black dots in the plot.
Their uncertainties are computed as per Appendix~\ref{sec:epoch_folding} and illustrate the better quality of the combined datasets analysis.
Note that in some cases the spin period estimates are not available from individual observations, however combining these datasets with adjacent ones often improves the pulse statistics.
The dashed gray line is a linear fit to the period estimates from combined datasets.
Dashed cyan line is the same but discarding the last period estimate due to its large uncertainty.
\label{fig:pdot}}
\end{figure}

This analysis of combined datasets reveals a spin-up trend observed in XB091D which is otherwise hard to be detected in individual observations, see Fig.~\ref{fig:pdot}.
Despite at least one probable $\dot{P}$ sign change (note e.g. a period increase in the next to last dataset comprising observations from Jun to Aug 2012), it is likely that the neutron star spins up.
On average the spin-up rate amounts to $\dot{P} \approx -5.7 \times 10^{-13}$~s~s$^{-1}$ if we consider all 9 period estimates obtained, or $\dot{P} \approx -7.1 \times 10^{-13}$~s~s$^{-1}$ if we reject the most recent dataset, having the largest period uncertainty.

\section{Spectral Analysis}
\label{sec:spec}
We considered all imaging observations in which XB091D was in the f.o.v. of the EPIC instruments. 
Data reduction was performed with the standard {\tt epproc}  and {\tt emproc} pipelines coming with the \xmm Science Analysis Software (SAS v.14) and using the latest calibration files.
From the raw event lists we selected high energy photons ($>$10 keV for the pn and between 10 and 12~keV for the MOS) and built light curves with 50~s binning, in order to identify and filter out time intervals affected by high particle background.
Threshold count rates were set at 0.4 $\rm ct\,s^{-1}$ and 0.35 $\rm ct\,s^{-1}$ for the pn and MOS, respectively.
We then used the interactive {\tt xmmselect} task from the SAS to extract source and background spectra from the 
filtered event lists, starting with ObsID 0690600401 which has the largest counts statistics. 

In this observation the source lies in the f.o.v. of all three EPIC cameras.
We extracted source counts from a circular region of 25\arcsec\ radius, 
while for the background we selected events from a nearby region free of sources within the same CCD. 
We applied the ({\tt FLAG==0})\&\&({\tt PATTERN==0}) filtering options during selection of pn events, in order to have the spectrum of the highest possible quality. 
For MOS counts we used the standard filtering flags, ({\tt \#XMMEA\_EM})\&\&({\tt PATTERN<=12}).
The extracted spectra were then re-binned in order to have at least 40 and 25 counts per energy bin for the pn and MOS, respectively.

\begin{figure}
\includegraphics[width=\linewidth]{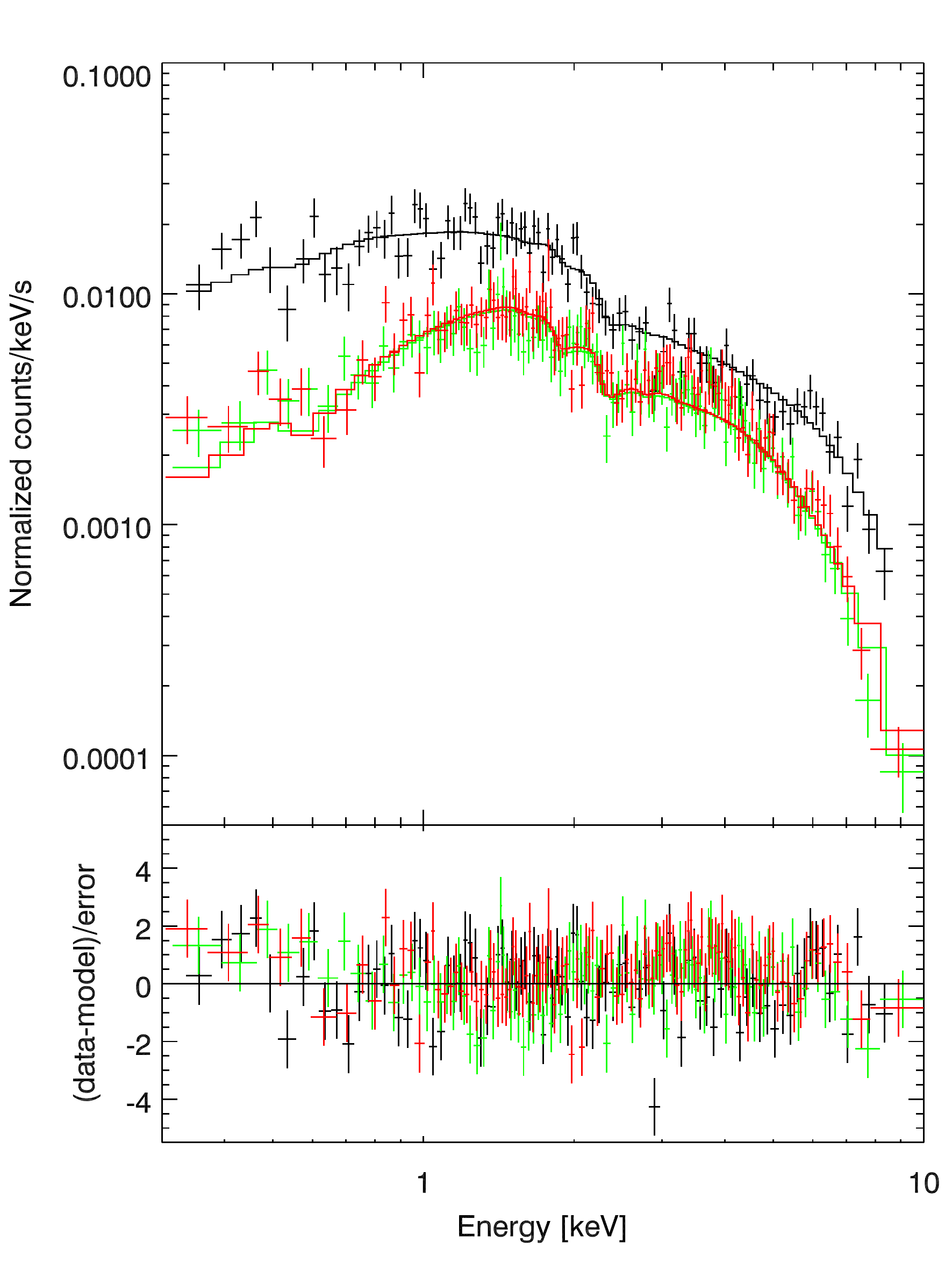}
\caption{Best-fit X-ray spectrum and folded model of XB091D during ObsID 0690600401.
pn data are black, MOS1 are green and MOS2 are red. See main text for details.
}
\label{fig:spectrum}
\end{figure}

Spectral fitting was performed with XSPEC v12.9 \citep{arnaud96}. 
The spectrum of XB091D can be described by an absorbed power law with exponential cut-off, {\tt wabs(cflux*cutoffpl)} in XSPEC.
The spectrum is hard, with a photon index $\Gamma=0.20\pm0.5$, a cut-off energy $E_{\rm cut}=4.6\pm0.4$~keV, and low absorption, $n_H = 3.79 \times 10^{20}$~\rm cm$^{-2}$, obtained from PN spectrum alone. We kept this value fixed in the simultaneous pn+MOS fit. This best-fit column density is in broad agreement with the expected value from the \citet{dl90} in the direction of \object{M31}. The reduced chi square of the fit is 1.16 for 310 degrees of freedom.
We present a plot of the best-fit folded spectrum and model for ObsID 0690600401 in Fig.~\ref{fig:spectrum}.

\begin{figure}
\includegraphics[width=\linewidth]{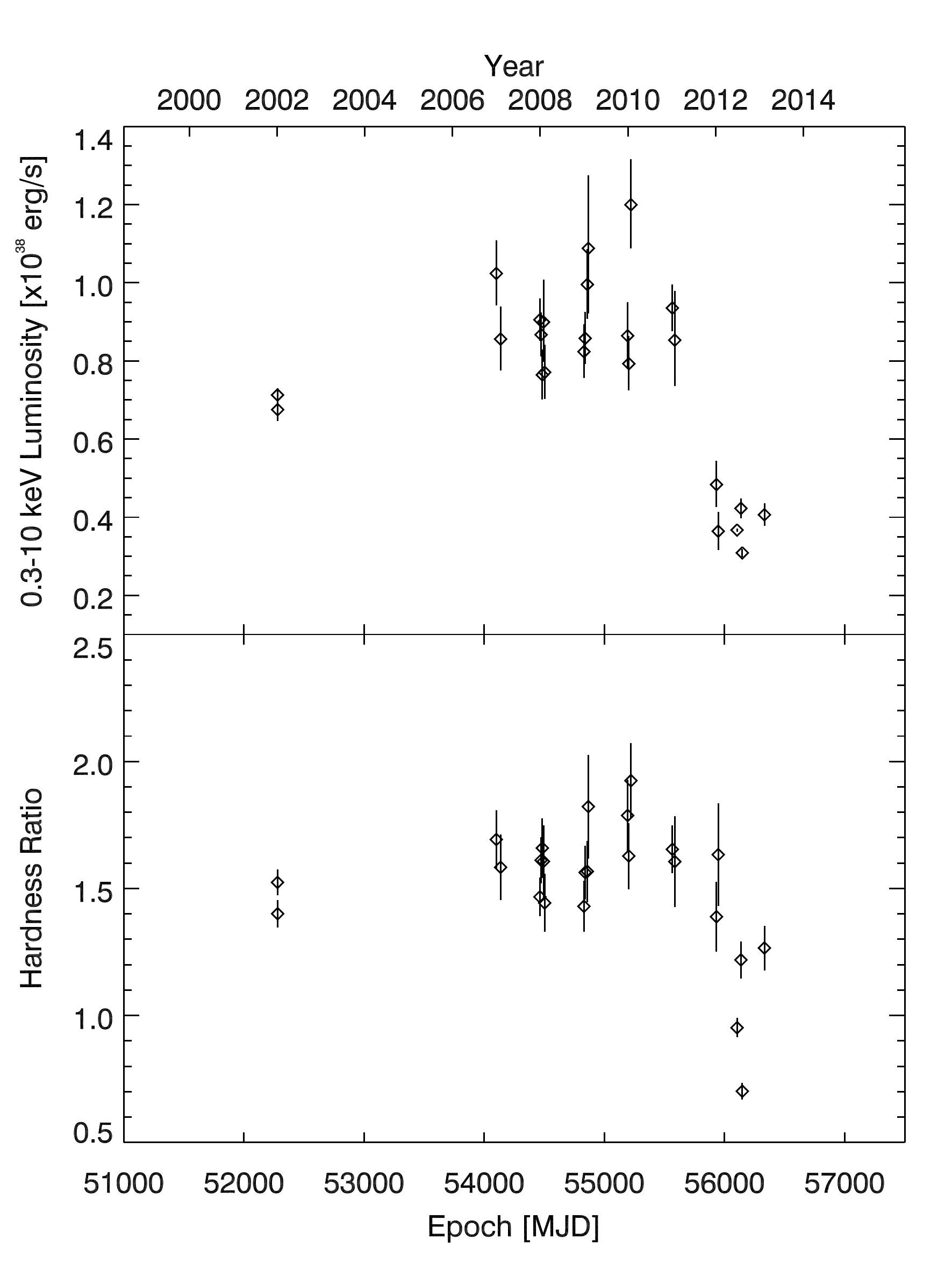}
\caption{${\it (Top)}$ Unabsorbed $0.3\textup{--}10$ keV luminosity of XB091D, in units of $10^{38}\,\rm erg/s$, versus the epoch of the observation.
${\it (Bottom)}$ Ratio of the unabsorbed fluxes estimated in the $5-10$ keV and $0.3-5$ keV bands, respectively. 
Error bars correspond to $1\sigma$ confidence limits.
}
\label{fig:spec_evo}
\end{figure}

\begin{figure}
\includegraphics[width=\linewidth]{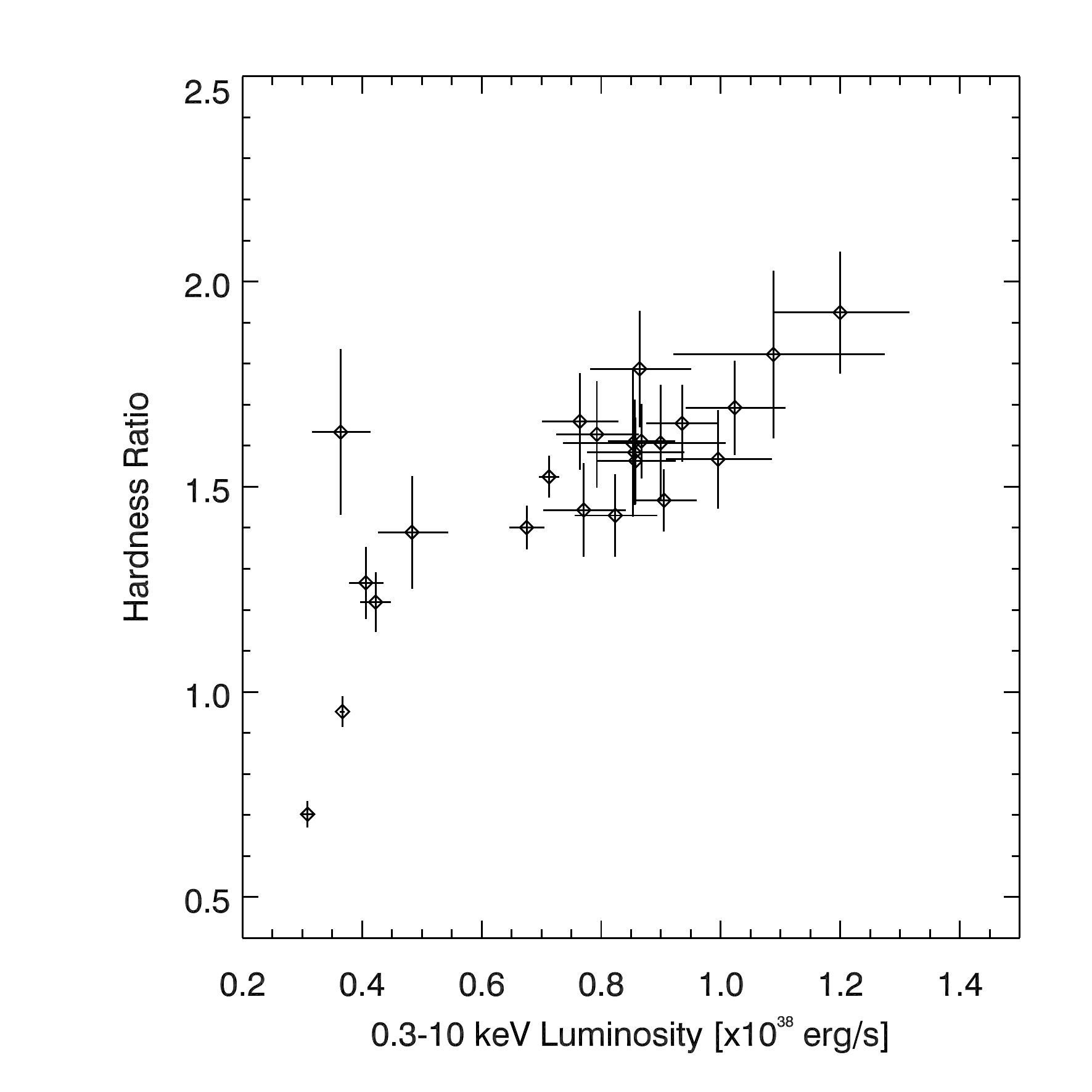}
\caption{Hardness ratio versus the unabsorbed $0.3\textup{--}10$ keV luminosity of XB091D. 
Error bars correspond to $1\sigma$ confidence limits.}
\label{fig:hr_vs_lum}
\end{figure}

We then analyzed all other observations, discarding those data sets where the source extraction region overlapped 
with CCD gaps or columns of bad pixels, and applying less stringent filtering options, ({\tt FLAG==0})\&\&({\tt PATTERN<=4}). 
In any case, given the lower count statistics, for all but ObsID 0700380601 a simpler absorbed power law model, 
{\tt wabs(cflux*powerlaw)}, is sufficient to fit the spectra adequately.
We used the multiplicative component {\tt cflux}, which returns the source's flux directly as a parameter of the fit, 
to characterize variations with the epoch of the unabsorbed $0.3-10$~keV band luminosity, and of the hardness ratio HR, 
where $\rm HR = Flux_{5-10\,keV}/Flux_{0.3-5\,keV}$, assuming a distance to \object{M31} of 752~kpc \citep{riess12} and isotropic emission (Fig.~\ref{fig:spec_evo}).
Intriguingly, the shape of the spectrum seems to be related with the luminosity of the source, where the harder spectra occur 
at higher luminosities, see Fig.~\ref{fig:hr_vs_lum}.

\citet{trudolyubov04} model the source X-ray spectrum with a hard power-law.
\citet{greening09} similarly find its photon index to be $0.9 \pm 0.1$ and interpret it as the high-mass X-ray binary (HMXB) nature of the source.
\citet{esposito16} note that for some observations the absorbed power-law model does not provide satisfactory results and fit a blackbody + power-law model, as well as a cut-off power-law, which agree with our results within the uncertainties. 
They also note a similar 'harder-brighter' correlation between the source luminosity and its spectral shape.

\section{Discussion}
\label{sec:discussion}

We first estimate the chance association probability of this X-ray source with the globular cluster B091D projection without being physically associated, as \citet{esposito16} favor an interpretation with the pulsar not associated with the GC. 
The association was first suggested by \citet{supper01} based on {\it ROSAT} data and confirmed by \citet{trudolyubov04} when the first {\it XMM-Newton} and {\it Chandra} data became available.
These works, however, do not provide an assessment of the probability of the false association.

XB091D is included in 3XMM-DR5 with R.A.=00:43:01.478, Dec=+41:30:16.94 (J2000) and 1$\sigma$ positional uncertainty of 0.13\arcsec. 
These coordinates are computed as a weighted average of all individual detections, most of them being largely off-axis.
The coordinates from ObsID 0690600401, the one with the smallest off-axis angle (1.82\arcmin), are: R.A.=00:43:01.440, Dec=+41:30:17.30 (J2000), or 0.56\arcsec\ away from the averaged coordinates, with a 1$\sigma$ positional uncertainty of 0.19\arcsec.
{\it Chandra} Source Catalog Release 1.1 \citep{evans10} includes this source as CXO J004301.4+413016 with coordinates R.A.=00:43:01.469, Dec=+41:30:16.80 (J2000) and the major semi-axis of the 95\% confidence level error ellipse of 1.19\arcsec.
The Bologna catalog gives B091D center coordinates as R.A.=00:43:01.446, Dec=+41:30:17.15 (J2000), which is consistent with the cluster center we find from the {\it HST} image within 0.15\arcsec\ of its astrometric calibration uncertainty.
Distances between the cluster center and the X-ray source coordinates are 0.41\arcsec\ for the averaged {\it XMM-Newton} position, 0.17\arcsec\ for ObsID 0690600401 position, and 0.44\arcsec\ for the {\it Chandra} position.
In our association probability calculations, we therefore assume that the X-ray source is located within 1\arcsec\ from the cluster center.

There are 38 GCs (including candidates) between 14 and 16\arcmin\ projected distance from \object{M31}'s center in the Bologna catalog which yields their density $6.0 \times 10^{-5}$ GCs per square arcsec at this angular distance from galaxy center. 
This gives a low probability ($1.2 \times 10^{-2}$) of having a GC within 1\arcsec\ of one of the 70 brightest X-ray sources from the 3XMM-DR5 catalog in \object{M31}. 
In fact, only 4 of the 70 brightest X-ray sources lie in this annulus. 
Therefore, their density of $6.3 \times 10^{-6}$ per square arcsec is translated to the probability of only $7.5 \times 10^{-4}$ of having a bright X-ray source within 1\arcsec\ of any of the globular clusters. 
It is therefore highly likely that this X-ray source belongs to B091D globular cluster.
Given the probability of coincidence, we assume below that the source belongs to the GC, and do not discuss alternative interpretations.

The detection of pulsations at $\sim0.83$ Hz secures the identification of the source with a spinning neutron star.
This is the first object of this class identified in the Andromeda Galaxy, and one of the most distant pulsars observed to date.
Also given that XB091D resides inside the globular cluster it spins slower than any known sources in a GC (see Introduction) and about ten times slower than the slowest accreting pulsar in a globular cluster known previously, the $\sim11$ Hz IGR~J17480-2466 in Terzan 5 \citep{papitto11}.
Its orbital period of 30.5~hr is also the longest known for accreting globular cluster binaries \citep[see table 5 in][]{bahramian14}.

Given the maximum observed X-ray luminosity of the system $L_{\rm X} = 1.2 \times 10^{38} $\,\ergs (see Fig.~\ref{fig:spec_evo}) which is close to Eddington, one can determine the corresponding mass accretion rate assuming that bolometric luminosity does not greatly exceed the X-ray luminosity: 
$\dot{M} = \frac{L_{\rm X} R_{\rm NS}}{G M_{\rm NS}} = 1.0 \times 10^{-8}$~\msun yr$^{-1}$.
We note that the observed X-ray variability of a factor $\simeq 3-4$ (see Fig.~\ref{fig:spec_evo}) excludes significant contribution from other unresolved X-ray sources residing in the same host globular cluster.
We cannot completely rule out the possibility that we observe 2 superimposed X-ray sources residing in the same globular cluster, though the probability of this coincidence is relatively low: in our Galaxy there is only one globular cluster known to host two bright persistent LMXBs, M15 \citep{white01}.
In case of the superposition we can only refer to the pulsed part of the flux (typically 20~per~cent, see Table~\ref{tab:obs}) as originating from this X-ray source.
This would adjust our calculation of the accretion rate and the magnetic field below by a small factor, without affecting the important conclusions.

For a neutron star spinning with a 1.2~s period, the corotation radius is 
$R_{\rm C} = (G M P^2 / 4 \pi^2 )^{1/3} \simeq 1890 M_{1.4}^{1/3}$~km, where $M_{1.4}$ is the NS mass in units of $1.4$\msun.

We employ same approach as in \citet{papitto11} to estimate the lower and upper limits for the magnetic field of the neutron star given that it persistently accretes matter from the companion.
It is reasonable to assume in this case that the accretion disk is truncated at the radius $R_{in}$ between the radius of the neutron star $R_{\rm NS}$ and the corotation radius $R_{\rm C}$: $R_{\rm NS} < R_{in} \lesssim R_{\rm C} = 1890 M_{1.4}^{1/3}$.
The inner disk radius can be approximately defined \citep[e.g.][]{GhoshLamb} from the balance between kinetic energy of the falling gas and the magnetic energy of the NS magnetosphere: $R_{in} \simeq 160 M^{1/7}_{1.4} R_6^{-2/7} L_{37}^{-2/7} \mu_{28}^{4/7}$, where $R_6$ is the neutron star radius in units of 10~km, $L_{37}$ is the accretion luminosity in units of $10^{37}$~erg~s$^{-1}$ and $\mu_{28}$ is the magnetic dipole moment of the NS in units of $10^{28}$~G~cm$^3$.
After trivial transformations we obtain: $0.008 L^{1/2}_{37_{\rm max}} < \mu_{28} \lesssim 75 L^{1/2}_{37_{\rm min}}$
We substitute the minimum ($L_{37_{\rm min}} = 3.5$) and maximum ($L_{37_{\rm max}} = 12.0$) observed X-ray luminosity into the resulting expression and obtain: $2.7 \times 10^8\,{\rm G} < B \lesssim 1.4 \times 10^{12}\,{\rm G}$.
This upper limit corresponds to the case of equilibrium rotation of the neutron star when the accretion disk is truncated at corotation radius and the falling matter does not transfer any angular momentum to the neutron star.
If the system is indeed in equilibrium, the derivative of the NS spin period $\dot{P}$ would be fluctuating in sign with small typical values. 
This is indeed what happens if one analyses individual observations of XB091D (see Fig.~\ref{fig:pdot}, light gray).
Assuming coherent rotation between adjacent observations separated by few months we were able to significantly improve the statistics and constrain period to much better precision which allowed us to detect constant average spin-up of the neutron star.

We find an average spin-up rate $\dot{\nu} \approx 4.0\textup{--}5.0 \times 10^{-13}$~Hz~s$^{-1}$, only a factor of two--three less than that of IGR~J17480$-$2446 which possesses a spin frequency derivative of $\dot{\nu} \approx 1.2 \times 10^{-12}$~Hz~s$^{-1}$ \citep{cavecchi11,patruno12}.
Such a significant average spin-up rate, if confirmed in future observations, may indicate that XB091D is not in equilibrium rotation and in this case its magnetic field estimate should be lower than the equilibrium value of $B \simeq 1.4 \times 10^{12}$~G.
IGR~J17480$-$2446's average spin-up rate agrees well with the accretion of disk matter angular momentum given the observed luminosity $2\textup{--}7 \times 10^{37}$\,\ergs and independent estimates of the inner radius of accretion disk from the quasi-periodic oscillations \citep{papitto11}.
From simple considerations a torque exerted on the neutron star by accreted material moving in a Keplerian disk is $N = 2 \pi I \dot{\nu} = \dot{M} \sqrt{G M R_{in}}$, where $I$ is the NS moment of inertia usually assumed to be $10^{45}$~g~cm$^2$.
Therefore we could roughly estimate the inner disk radius to be $R_{in} \simeq 30$~km and corresponding magnetic field $B \simeq 5 \times 10^8$~G computed for the minimum observed accretion rate $\dot{M} = 2.6 \times 10^{-9}$~\msun~yr$^{-1}$.
This simple computation yields results which are very similar to the ones for IGR~J17480$-$2446: $R_in \lesssim 20$~km, $B \simeq 7 \times 10^8$~G \citep{papitto11}. The observed average spin-up rate is not compatible with this simple model for the maximum observed accretion rate $\dot{M} = 1.0 \times 10^{-8}$~\msun~yr$^{-1}$ as it gives $R_{in}$ below 10~km.

It is not clear to what extent this $\dot{P}$ estimate is influenced by the pulsar's acceleration along the line of sight $a_l$ in the globular cluster gravitational field, which adds a term $a_l / c$ to $\dot{P} / P$.
However, the largest known absolute value of $\dot{P} / P$ from pulsars in globular clusters (measured in B1718$-$19 in NGC~6342) amounts to $\dot{P} / P \approx 1 \times 10^{-15}$~s$^{-1}$ and can be used as the upper limit of $a_l / c$ for pulsars in globular clusters.
We therefore do not expect it to influence our spin-up rate estimate much.

On the other hand, many accreting X-ray pulsars (including LMXBs) are known to undergo lengthy episodes of spin-ups and spin-downs, at the same time staying close to their equilibrium rotation period \citep[e.g.][]{chakrabarty97,bildsten97,perna06}.
If confirmed, the penultimate dataset in Fig.~\ref{fig:pdot} indicates at least one switch from the spin-up to the spin-down regime.
We therefore cannot exclude the equilibrium rotation of the neutron star in XB091D and hence larger value of its magnetic field $B \simeq 1.4 \times 10^{12}$~G.

XB091D exhibits rather small luminosity changes during the observed 11 years span (see Fig.~\ref{fig:spec_evo}) which increases to 12 years if we consider an ObsID 0112570601 from the end of 2000 at very large off-axis angle which however shows very similar flux in the 3XMM-DR5 catalog.
The hard X-ray spectrum (see Fig.~\ref{fig:spectrum}) of the source is typical for LMXBs, however it cannot be used alone to discard other possible interpretations such as an HMXB nature.
Keeping in mind the association with a globular cluster, this evidence allows us to classify the source as a persistent LMXB.
Assuming 100~per~cent duty cycle and persistent accretion at the observed rate, 
it becomes possible to estimate time to spin up XB091D to millisecond period, $P / \dot{P} \simeq 50\textup{--}100 \times 10^3$~yr.
Extrapolating to the past, it would take XB091D only 1~Myr or even less to spin up from 10~s period to its observed value.

\subsection{Properties of host globular cluster B091D and comparison with Galactic globular clusters}
\label{sec:gccomp}

In order to compare B091D to Terzan~5, the host cluster of the slowest known
accreting X-ray pulsar in a globular cluster \citep{papitto11}, and several other massive
globular clusters in the Milky Way hosting dense stellar cores, we used archival imaging and
measurements of stellar velocity dispersion and stellar populations of
globular clusters from literature.
Knowing the cluster structural parameters (e.g. 
the central stellar density $\rho_0$ and the core radius $r_c$) and central
stellar velocity dispersions $\sigma_0$ in cluster cores, we can directly
compare the numbers of LMXB formation events by tidal capture using the
relation for encounters rate from \citet{verbunt87}: 
\begin{equation}
\Gamma \propto \rho_0^2 r_c^3 / \sigma_0
\label{lmxb_cap_eq}
\end{equation}

Terzan~5 is a massive ($M = 2 \times 10^6$~\msun) very metal rich cluster
([Fe/H]~$\simeq 0$~dex) hosting a rich population of X-ray binaries. 
It possesses multiple stellar populations \citep{Ferraro+09} so it is
believed to be a nucleus of a dwarf galaxy heavily stripped by the Milky
Way.
It has a small core radius $r_c = 0.16$~pc, a very high central stellar density $\rho_0 = 1.58 \times 10^6$~\msun~pc$^{-3}$ \citep[][from the long-term timing observations of pulsars population]{prager16}, and a tidal radius $r_t \approx 6.7$~pc \citep{Lanzoni+10}.
Its central velocity dispersion is estimated to be $\sigma_0 = 12.7$~km~s$^{-1}$ \citep{Gnedin+02}.  
Terzan~5 has one of the densest known stellar cores among globular clusters in our Galaxy \citep{cohn02}.

B091D was included in the sample of \object{M31} GCs with stellar velocity dispersions measured from high resolution optical spectra \citep{SCS11}. 
However, their dynamical models are based on ground based images and, therefore, might be inaccurate.

We downloaded high resolution Hubble Space Telescope optical images for
B091D from the Barbara A.  Mikulski Archive for Space Telescopes\footnote{\url{http://mast.stsci.edi/}} obtained in the framework of the
HST GO program 10273 ``Accurately Mapping M31's Microlensing Population'' (P.I.: A.~Crotts).  
The cluster is located close to the edge of the field of view on two single exposures in the \emph{F555W} (exposure time $t_{exp}=151$~sec) and \emph{F814W} ($t_{exp}=457$~sec) filters obtained with the HST Advanced Camera for Surveys Wide Field Camera.  
Several pixels in the $F814W$ image close to the cluster core are saturated.  
We generated HST point-spread functions in the two filters using the {\sc tinytim} software \citep{KHS11} and then ran the {\sc galfit} 2-dimensional image
fitting code \citep{PHIR02} and fitted King (\citeyear{King66}) profiles into
B091D images masking saturated pixels in the $F814W$ image.  

We obtained the following parameters for the $F555W$ profile: core radius
$r_c = 0.110$\arcsec$\,= 0.42$~pc, truncation radius $r_t = 7.1$\arcsec$\,= 27$~pc, central surface brightness $\mu_{0,555}=13.61$~mag~arcsec$^{-2} $ or $I_0 = 1.31\times10^5 L_{\odot}$~pc$^{-2}$ that corresponds to $\rho_0
\approx 8\times10^5$~\msun~pc$^{-3}$, ellipticity $e=0.92$. 
Uncertainties on the structural parameters are an order of 5--7~per~cent. 
The parameters obtained from fitting the $F814W$ image ($r_c = 0.15$\arcsec,
$r_t = 6.3$\arcsec, $\mu_{0,814}=13.44$~mag~arcsec$^{-2}$, $e=0.89$) are
consistent within uncertainties with those obtained from $F555W$ except the
core radius probably affected by saturated pixels.  
We notice that \citet{agar13} used the same datasets and performed similar
analysis. However, for some reason, their values of central surface
brightness and core radius derived from the $F555W$ image differ from the
$F814W$ one by almost three orders of magnitude and look quite unrealistic.
The latter set of parameters ($F814W$) agrees reasonably well with our estimates.
We note, that our $r_c$ value is somewhat smaller than a core radius
$r_0 = 0.54$~pc reported by \citet{peacock10} obtained from the analysis of
ground based near-infrared $K$-band images.

Following \citet{RT86} and converting into proper units, we estimate the $V$
band dynamical mass-to-light ratio of B091D as: $(M/L)_{V} \approx 333
\sigma_0^2 / (r_c I_0)$, where $\sigma_0$ is a central projected stellar
velocity dispersion in km~s$^{-1}$, $r_c$ is a core radius in pc, $I_0$ is a
central surface brightness in $L_{\odot}$~pc$^{-2}$.  
The aperture correction for the observed value $\sigma=18.6 \pm 1.0$~km~s$^{-1}$ \citep{SCS11} obtained by the integration of the King model yields $\sigma_0=21.0\pm1.3$~km~s$^{-1}$, hence $(M/L)_{V,\mbox{dyn}} = 2.6\pm0.4 (M_{\odot}/L_{\odot})_{V}$ or $M_{dyn} = (9.6 \pm 1.5) \times 10^5$~\msun assuming $V=15.39$~mag.

Substituting these values in Eq.~\ref{lmxb_cap_eq}, we estimate the ratio of
the stellar encounter rates in Terzan~5 and B091D to be:
$\Gamma_{\mathrm{Terzan~5}}/\Gamma_{\mathrm{B091D}} \simeq 0.34$ that means that the LMXB formation by capture in the core of B091D is 2.9 times more likely than in Terzan~5, one of the densest globular clusters in our Galaxy which also possesses the richest population of X-ray sources observed in a globular cluster \citep{heinke06}.
As encounter number $\Gamma$ was shown to correlate with the numbers of X-ray binaries in globular clusters \citep{pooley03} we can therefore expect B091D to be quite prolific globular cluster having X-ray binaries population similar or larger than that of Terzan~5 and continuously forming new systems at present epoch.

\citet{Caldwell+11} reported old stellar population ($t\approx12$~Gyr) and a metallicity ${\rm [Fe/H]} = -0.70$~dex for B091D making it a representative of ``red'' metal-rich globular clusters.  
For the \citet{Kroupa02} stellar initial mass function, these parameters correspond to the stellar mass-to-light ratio $(M/L)_{*,V} = 2.6 (M_{\odot}/L_{\odot})$.  
This remarkable agreement with the dynamical $(M/L)_{V,\mbox{dyn}}$ suggests that B091D did not experience stellar mass loss due to tidal stripping by \object{M31}
after dynamical relaxation and stellar mass segregation if it had been born
with the Kroupa IMF.  

Assuming a simple spherically symmetric model, the estimated two-body relaxation timescale of B091D is $t_{\mbox{relax}} \approx 1$~Gyr
\citep{spitzer71,spitzer87}. 
Therefore, it is not expected to have undergone the core collapse  because it occurs at $t_{cc} \approx 17 t_{\mbox{relax}}$ \citep{BHH02}. 
The dense stellar core in B091D probably formed over 10~Gyr ago and is not a result of its secular evolution.

We use the recent dynamical analysis for a large sample of Galactic globular
clusters by \citet{Baumgardt17} and additional data from the Harris catalog
\citep{Harris96}, 2010 edition\footnote{Electronic version available online at:
\url{http://www.physics.mcmaster.ca/~harris/mwgc.dat}} to compare B091D and similar clusters from the Milky Way (see Table~\ref{tab:gcs}). 
It turns out that none of the Galactic globular clusters reaches the central velocity dispersion $\sigma_0=21$~km~s$^{-1}$ observed in B091D. 
The two closest analogs are NGC~6388 and NGC~6441.
Interestingly, similarly to B091D they also both harbor metal rich stars ($-0.55$ and $-0.46$~dex), are massive (masses slightly over $10^6$~\msun), and possess dense stellar cores with central densities around $4\textup{--}5 \times 10^5$~\msun pc$^{-3}$, which were probably reached primordially because both clusters have expected core collapse times exceeding the Hubble time. 
Two other clusters, although having much lower stellar metallicities, with central velocity dispersions approaching 20~km~s$^{-1}$ analyzed by \citet{Baumgardt17} are $\omega$~Cen and \object{M54}.  
The latter one also posseses a compact dense stellar core similar to B091D.
A common characteristic of these four objects and B091D is their high dynamical mass-to-light ratios, which suggest that their present day stellar mass functions are compatible with the Kroupa IMF for B091D, NGC~6388 and NGC~6441 and is even slightly steeper at low masses for \object{M54} and $\omega$~Cen.  
Conversely, the vast majority of other Galactic globular clusters are substantially less massive. They, therefore, have undergone the depletion of low-mass stars by the tidal stripping because they are globally mass segregated so that low-mass stars migrated to the outskirts, hence their dynamical mass-to-light ratios have lowered compared to what one would expect for the Kroupa IMF\footnote{It should be understood, the compact cores themselves have much shorter two-body relaxation times, typically tens of Myr, and they are probably
mass segregated and dynamicaly evolved as demonstrated by \citet{Heyl+15} 
based on a population of young white dwarfs in 47~Tuc.}.

It is agreed in the community, that well studied $\omega$~Cen \citep{bekki03,Gnedin+02} and \object{M54} \citep{siegel07} are likely the tidally stripped nuclei of dwarf galaxies; \object{M54} is thought to be residing at the core of the Sagittarius dwarf spheroidal galaxy.
This makes them analogous to ultra-compact dwarf (UCDs) galaxies \citep{Drinkwater+03} observed in the nearby Virgo and Fornax clusters, which are known to have mass-to-light ratios consistent with the Kroupa IMF when the two-body relaxation timescales are long \citep{CMHI11}.  
UCDs are proven to be tidally stripped galactic nuclei by the recent discovery of central massive black holes \citep{Seth+14} and by their stellar population properties \citep{Francis+12} similar to much more massive tidally stripped M32-like compact elliptical galaxies \citep{Chilingarian+09,CZ15}.
The only known stellar systems where central luminosity densities exceed $10^5$~\lsun pc$^{-3}$ are nuclear star clusters in low and intermediate luminosity galaxies, cEs, some UCDs \citep{evstigneeva08} and some globular clusters.
Among $\sim$160 Galactic globular clusters in the Harris catalog, only a handful possess $\rho_0>10^5$~\lsun pc$^{-3}$ and are not marked as core collapsed.
Three of them are Terzan~5, NGC~6388, and NGC~6441. 
In the light of new structural and dynamical data (namely $\sigma_0$, $\rho_0$, M/L ratios) presented in \citet{Baumgardt17} and also keeping in mind that stripped nuclei should contribute to the high end of GC luminosity function \citep[see e.g.][]{pfeffer14}, we suspect that NGC~6388 and NGC~6441 are, in fact, heavily tidally stripped galactic nuclei.
Interestingly, these two clusters have extremely extended and peculiar horizontal branches making them unique compared to other GCs \citep{brown16}, which can be explained only by very high helium abundances \citep[see][]{bellini13,brown16} that may be an evidence of self-enrichment during the extended period of star formation common for galactic nuclei but uncommon for globular clusters.
So as B091D, given how similar they are in terms of both stellar population and internal structure and dynamics. 
They, however, must have undergone a somewhat higher degree of tidal stripping than \object{M54}.

Keeping in mind that Terzan~5 hosting another slow X-ray pulsar is also suspected to be a stripped galaxy nucleus, this raises a question whether some specific conditions in nuclear star clusters favor the formation of LMXBs.  
Possibly, very high stellar densities reached in galactic nuclei similar to those in core collapsed globular clusters but on an order of magnitude larger spatial scale (see table~2 in \citealp{evstigneeva08} for central surface brightness and size values of UCD cores) provide an effective formation channel of relativistic binaries via close encounters (see also \citet{DKPM12} on overabundance of LMXBs in UCDs).
We illustrate this by comparing $\Gamma$ values of discussed clusters in Table 3.
NGC~6388, NGC~6441 and B091D have large central densities $\rho_0$ and at the same time large cores $r_c$, and therefore very high encounter rates $\Gamma$ which even exceed the one of the densest cluster Terzan 5.

We also compare another dynamical parameter for GCs, the encounter rate for a single binary, $\gamma \propto \rho_0 / \sigma_0 $ \citep{verbunt03,verbunt14}.
It is expected that a higher $\gamma$ indicates a higher rate of exchange encounters in a globular cluster and therefore higher observed frequency of exchange encounter products -- such as isolated pulsars, slow young neutron stars and other kinds of exotic objects believed to be formed by the disruption of X-ray binaries.
The lifetime of a binary until the next encounter which increases their chances to get disrupted or to exchange companion star is proportional to $1 / \gamma$.
We computed $\gamma$ values for B091D, Terzan~5, and the four
clusters discussed above which we think can all be classified as UCDs using
the updated global mass-to-light ratios and central velocity dispersions from
\citet{Baumgardt17} and structural parameters from the Harris catalog. We
note, that \citet{Baumgardt17} demonstrated (see their fig.~3) that the
M/L ratio in the cluster core is close to the global value while the radial
M/L profiles often exhibit a dip explained by the mass segregation.
In the units of the reference globular cluster M4 from \citet{verbunt14}, $\gamma_{\rm B091D} = 23 \, \gamma_{\rm M4}$.
This makes B091D similar to the top 5 GCs of our Galaxy by this parameter.

\begin{deluxetable}{lllllll}
\tablecolumns{7}
\tablecaption{Structural parameters of B091D and similar globular clusters in the Milky Way, their stellar encounter rate $\Gamma$, and their stellar encounter rate for a single binary $\gamma$.}
\tablehead{   
  \colhead{Name} &
  \colhead{log~$\rho_0$\tablenotemark{a}} &
  \colhead{M/L\tablenotemark{b}} &
  \colhead{$r_c$\tablenotemark{a}} &
  \colhead{$\sigma_0$\tablenotemark{b}} &
  \colhead{$\Gamma$} &
  \colhead{$\gamma$} \\
  \colhead{} &
  \colhead{($L_{\odot}$/pc$^{3}$)} &
  \colhead{($\odot$)} &
  \colhead{(pc)} &
  \colhead{(km/s)} &
  \colhead{} &
  \colhead{}
}
\startdata
\textit{M4} & \textit{3.64} & \textit{1.70} & \textit{0.72} & \textit{4.5} & \textit{1} & \textit{1} \\
Terzan~5          & 5.78\tablenotemark{c} & 2.6\tablenotemark{c} & 0.16\tablenotemark{c} & 13\tablenotemark{d} & 170 & 73 \\
NGC6388       & 5.37 & 2.11 & 0.38 & 15 & 200 & 20 \\
$\omega$~Cen & 3.15 & 2.54 & 3.45 & 17 & 6.7 & 0.12 \\
NGC6441      & 5.26 & 2.30 & 0.51 & 17 & 300 & 15 \\
M54                & 4.69 & 2.18 & 0.62 & 19 & 30 & 3.4 \\
\textbf{B091D} & \textbf{5.49}\tablenotemark{e} & \textbf{2.6}\tablenotemark{e} & \textbf{0.42}\tablenotemark{e} & \textbf{21}\tablenotemark{e} & \textbf{490} & \textbf{23}
\enddata
\tablenotetext{a}{\citet{Harris96}, 2010 edition, unless noted}
\tablenotetext{b}{\citet{Baumgardt17}, unless noted}
\tablenotetext{c}{\citet{prager16}, M/L in the center}
\tablenotetext{d}{\citet{Gnedin+02} King model estimate assuming $M/L$=3}
\tablenotetext{e}{This study}
\tablecomments{$\omega$~Cen does not have a compact core but is included here because of its high central velocity dispersion.
M4 is included only as the $\Gamma$ and $\gamma$ unit scale.
$\Gamma$ is computed as $(\rho_0 M/L)^2 r_c^3 / \sigma_0$, and $\gamma$ is computed as $(\rho_0 M/L) / \sigma_0$, where $\rho_0$ is a central volume luminosity density; result is rounded to 2 significant figures.
The list is sorted by the central velocity dispersion.
\label{tab:gcs}}
\end{deluxetable}

\subsection{System age and formation scenarios}
\label{sec:age}

\citet{Caldwell+11} estimated the age of B091D globular cluster to be 12~Gyr.
Generally, there exists few evolutionary sequences of binaries formed more than 10~Gyr ago that start an accretion episode of required intensity at the present time and hence could explain the origin of XB091D.
For instance, in \citet{podsiadlowski02} binaries with initial mass of a secondary star $M_2$ between 1.0 and 1.2 \msun, initial orbital period $P$ between 0.5 and 
100 days, exhibit accretion episodes of order $\simeq 100$~Myr in duration after $\simeq 10\textup{--}12$~Gyr of evolution, reaching peak accretion rate 
$\dot{M}_{\rm peak} \simeq {\rm few} \times 10^{-8}$~\msun yr$^{-1}$ with average accretion rate $\dot{\left<M\right>} \simeq {\rm few} \times 10^{-9}$~\msun yr$^{-1}$.
This shows that scenario of primordial origin of XB091D when the system formed at early epochs of its host globular cluster around 12~Gyr ago and started accretion episode which we observe today very recently, is not forbidden by the evolution theories of isolated binary systems.

A scenario when XB091D hosts a primordial neutron star must explain how a neutron star which formed 12~Gyr ago kept the most probable value of its current magnetic field $B \simeq 1.4 \times 10^{12}$~G, provided that neutron stars are thought to be born with magnetic fields $B \simeq 10^{13}\textup{--}10^{14}$~G \citep{faucher06}.
It is not clear whether non-accreting neutron stars preserve their magnetic field for a long time or not. 
Some models estimate it to decay with the characteristic time that spans from $\sim 10^6$ \citep[e.g.][]{narayan90} to $\sim 10^8$ years \citep[e.g.][]{bhattacharya92}.
Given the lack of consensus about the magnetic field decay in neutron stars, we cannot rule out the primordial neutron star origin in XB091D.
In this context, we note that \citet{ivanova08} find in their simulations that less than 10\% of the primordial binaries in a given GC that survived core collapse supernova, remain in the original system after 11~Gyr (see their table 4).
In this case if XB091D hosts a primordial neutron star, it must have acquired the current secondary at a later epoch.
Then a probable channel to get a companion is a binary exchange \citep{verbunt14}.
This is supported in particular by the fact that exchange encounters favor wide binary systems like XB091D with orbital periods of more than 1 day \citep[e.g.][]{verbunt03}.

Alternative formation scenario that is capable of producing a neutron star in an old globular cluster is an accretion-induced collapse (AIC) when a massive ($\simeq 1.2 M$) ONeMg white dwarf (WD) accretes matter from a companion until it reaches the Chandrasekhar limit $M = 1.44$\msun.
AIC is anticipated to be responsible for the population of slow isolated pulsars with high magnetic fields in GCs \citep[e.g.][]{lyne96,breton07} and to be the origin of some slow accreting X-ray pulsars in the field such as 4U~1626$-$67 \citep[e.g.][]{yungelson02}, though it has never been observed directly.
\citet{ivanova08} claim that in a typical globular cluster during 9.5$-$12.5~Gyr production of LMXBs from AIC is two to three times more efficient than any other dynamical formation channel, such as physical collisions, tidal captures and binary exchanges.
A cluster like Terzan~5 (and therefore very similar B091D) is thought to produce from AIC $9.35 \pm 1.20$ LMXBs per Gyr at ages $11 \pm 1.5$~Gyr \citep{ivanova08} which is not negligible even considering the short lifetime of such binaries.
One characteristic property of a NS formed in AIC event is its low mass, $M_{\rm NS} \simeq 1.26$~\msun.
We doubt however that it is possible to constrain the neutron star mass in this system to support or discard an AIC hypothesis.
No optical identification and spectroscopic observations of companion star is possible with current generation of astronomical instrumentation of a source in globular cluster at \object{M31} distance.
In case of AIC origin of the neutron star in XB091D there are several reasons to suspect that we currently observe accretion episode powered by a new donor after dynamical exchange event took place with the original binary that hosted AIC.
During AIC white dwarf loses roughly 0.2\msun in the form of binding energy and probably some mass in a supernova shell, which makes the binary orbit wider therefore detaching binary and halting mass transfer.
Time between AIC and resumption of mass transfer in the ultra-compact system with white dwarf donor is $\simeq 10^8$~yr \citep{verbunt90}, though it obviously strongly depends on donor properties -- for example on the presence of magnetic braking that brings the secondary into contact with its Roche lobe, or the rate of the secondary star radius increase due to its nuclear evolution.
On the other hand, the donor loses a significant fraction of its mass to power AIC event so that it can be incapable of powering an intensive accretion episode, which we are currently observing.
Prolonged epochs without accretion increase the chances for the binary exchange inside a globular cluster with a high specific encounter rate $\gamma$.

The mass loss of the original donor to trigger the AIC event should be within $0.2\textup{--}0.3$~\msun range.
From our orbital solution we estimate the donor mass function to be 0.0160, in agreement with \citet{esposito16}.
This is translated to the minimum donor mass of $M_2 = 0.36$\msun in case of edge-on system with inclination angle $i = 90$~deg.
The lack of X-ray eclipses in the longest observation which covers all orbital period means we can constrain the donor mass a little further as the system's inclination is then less than $\simeq 70$~deg: $M_2 > 0.38$\msun.
For a random distribution of inclination angles, one has the 90~per~cent a priori probability of observing a binary system at an angle $i > 26$~deg.
For the observed mass function, this inclination corresponds to $M_2 = 1.04$\msun.
Therefore the 90~per~cent confidence interval for the donor mass is $0.38 \leq M_2 \leq 1.04$\msun.
In fact for a 12~Gyr old globular cluster B091D the main sequence turn-off mass is 0.8\msun.
All stars within B091D with mass $\gtrsim 1.0$~\msun must have evolved to red giants and even become white dwarfs.
So the reasonable upper limit on the donor mass is $M_2 \simeq 0.8\textup{--}0.9$~\msun which corresponds to the low-mass sub-giants and stars leaving the main sequence.

Using \citet{eggleton83} formula and considering limits for the mass ratio ($M_1 = 1.25$\msun, $M_2 = 0.9$\msun and $M_1 = 2.0$\msun, $M_2 = 0.4$\msun) it is easy to estimate the size of the Roche lobe for the companion star: $1.64 \leq R_{L_2} \leq 2.24$\rsun.
Therefore, to fill its Roche lobe the donor must be an evolved star which has recently left the main sequence and has its radius increased to about $2$\rsun.
The current high accretion rate can be understood as driven by the nuclear evolution of the binary with rapidly increasing radius.

It is very unlikely that a low-mass star with $M_2 \simeq 0.4\textup{--}0.5$\msun (e.g. donor that powered AIC event) can reach 1.6\rsun radius.
This could however be influenced by the X-ray irradiation of the donor surface but its effect is not well understood presently.
For the most likely value of the donor mass $M_2 = 0.8$\msun the inclination of the orbital plane is $\simeq 30$~deg and the orbital separation is 6.27--6.96\rsun (0.029--0.032 AU).
This makes XB091D the widest known accreting binary system in a globular cluster.

The pulsar recycling theory \citep{bhattacharya91} assumes that after rotation-powered phase of classical pulsar finishing with $P_{\rm spin} \approx 1\textup{--}10$~s neutron stars can be spun up to $P_{\rm spin} \approx 1\textup{--}10$~ms by an accreting donor in binary system.
XB091D very well fits into this picture being observed at the earliest stages of its accretion spin-up phase.
Whereas very similar pulsar IGR~J17480$-$2466 from Terzan~5 represents a mildly recycled system, XB091D is a missing example of non-recycled neutron star which nonetheless is accreting very intensively.
The endpoint of the evolution of XB091D in a few million years is likely to be a millisecond radio pulsar in a wide ($\simeq$ several days) orbit with a white dwarf companion. Such MSP systems are numerous, but their progenitors like IGR~J17480$-$2466 and XB091D have relatively short lifetimes at high mass transfer rates, so very few of them are observed.

\section{Conclusions}
\label{sec:conclusions}

We report an independent detection of a luminous ($L_{\rm X} = 3\textup{--}12 \times 10^{37}$\,\ergs) accreting X-ray pulsar in the Andromeda galaxy in the public data of 38 observations obtained by the {\it XMM-Newton} observatory between 2000 and 2013. 
In 13 observations we detected 15 to 30~per~cent pulsed emission with period of 1.2~s.
Our analysis is fully reproducible online using the \xmm photon database available at \url{http://xmm-catalog.irap.omp.eu}.

We demonstrate that this X-ray binary is associated with a massive 12~Gyr old globular cluster B091D hosting a very dense stellar core and possessing a high stellar encounter rate. 
Therefore this system is a very unusual example of a non-recycled pulsar intensively accreting.
At 1.2~s its neutron star spins 10 times slower than the former slowest known X-ray pulsar in globular clusters -- a mildly recycled system IGR~J17480$-$2446 in Terzan 5.

From the X-ray timing analysis we estimate the binary system orbital parameters, including its orbital period of 30.5~h.
By combining several adjacent datasets in order to increase the photon
statistics, we obtain a phase connected solution in 9 extended periods of time over the 11 years baseline and detect an average neutron star's spin-up rate of $\dot{P} \simeq -7.1 \times 10^{-13}$~s~s$^{-1}$, which however has at least one episode of spin-down.
If the system is not in an equilibrium rotation and the spin-up persists, from this number we estimate that the accretion onset happened less than 1~Myr ago, because it will only take $\approx 10^5$~yr for this system to become a conventional millisecond pulsar.

The detected average spin-up rate matches that expected from the angular momentum accretion of Keplerian disk with the accretion rate $\dot{M} \simeq 3 \times 10^{-9}$~\msun~yr$^{-1}$ if the inner boundary of the disk is at $R_{in} \simeq 30$~km.
If we assume that the neutron star in XB091D indeed is not in the equilibrium rotation, we can estimate its magnetic field to be $B \simeq 5 \times 10^8$~G. 
The observed change to spin-down, however, favors an equilibrium configuration with a larger value $B \simeq 1.4 \times 10^{12}$~G which is also supported by the hard observed X-ray spectrum.
From the orbital separation and the donor Roche lobe size, also keeping in mind that the system has been persistently accreting over the past 12 years, we conclude that the secondary must be a slightly evolved low-mass star with the mass close to the main sequence turn-off for a 12~Gyr old globular cluster $M_2 \simeq 0.8$\msun.

Based on these properties, we cannot distinguish between a primordial NS origin and its formation at a later epoch e.g. in the AIC event. 
However, in both cases it is highly unlikely that we observe an original binary where the NS was formed.
In the AIC case the system has likely experienced an exchange interaction and the neutron star captured a non-exhausted low-mass donor star.
In the case of a primordial NS it must have acquired a secondary star after it was formed.
In both cases the most likely process of getting a donor star is an exchange encounter and the donor started to overflow its Roche lobe very recently, less than 1~Myr ago.
These formation scenarios are in line with the measured properties of XB091D and correspond to the expectations that follow from the global properties of the host globular cluster, namely its high encounter rate for a single binary $\gamma$, a predicted indicator of the frequency of binary systems that form via exchange encounters.
XB091D is the first accreting non-recycled X-ray pulsar which completes the picture of pulsar recycling.

\section*{Acknowledgments} 
This work is based on observations obtained with {\it XMM-Newton}, an ESA science mission with instruments and contributions directly funded by ESA Member States and the USA (NASA).
This research has made use of the VizieR catalogue access tool, CDS, Strasbourg, France. 
Part of the plots were produced using Veusz by Jeremy Sanders.
Authors are grateful to citizen scientists M.~Chernyshov, A.~Sergeev, and A.~Timirgazin for their help with the development of the \xmm catalog website \url{http://xmm-catalog.irap.omp.eu} used throughout this study.
The authors thank N.~Ivanova for the useful comments on the paper and H.~Baumgardt for providing structural parameters for globular clusters and discussion regarding the globular cluster dynamics.
IZ acknowledges the support by the Russian Scientific Foundation grant 14-50-00043 for the data processing and grant 14-12-00146 for the timing analysis.
IC and IZ acknowledge the joint RFBR/CNRS grant 15-52-15050 supporting the Russian--French collaboration
on the archival and Virtual Observatory research, the RFBR grant 15-32-21062 and the president of the Russian Federation grant MD-7355.2015.2
supporting the studies of globular clusters and compact stellar systems. 
The work of NS was supported by the French Space Agency CNES through the CNRS. MB was supported by the Sardinian Region through a fundamental research grant under Regional Law 7th.
Part of the detection chain was adapted from the software library for X-ray timing MaLTPyNT \citep{bachetti15ascl}.

\bibliography{m31_pulsar,matteo}

\begin{thebibliography}{}
\expandafter\ifx\csname natexlab\endcsname\relax\def\natexlab#1{#1}\fi

\bibitem[{{Agar} \& {Barmby}(2013)}]{agar13}
{Agar}, J.~R.~R., \& {Barmby}, P. 2013, \aj, 146, 135

\bibitem[{{Alpar} {et~al.}(1982){Alpar}, {Cheng}, {Ruderman}, \&
  {Shaham}}]{alpa82}
{Alpar}, M.~A., {Cheng}, A.~F., {Ruderman}, M.~A., \& {Shaham}, J. 1982, \nat,
  300, 728

\bibitem[{{Archibald} {et~al.}(2009){Archibald}, {Stairs}, {Ransom}, {Kaspi},
  {Kondratiev}, {Lorimer}, {McLaughlin}, {Boyles}, {Hessels}, {Lynch}, {van
  Leeuwen}, {Roberts}, {Jenet}, {Champion}, {Rosen}, {Barlow}, {Dunlap}, \&
  {Remillard}}]{arch09}
{Archibald}, A.~M., {Stairs}, I.~H., {Ransom}, S.~M., {et~al.} 2009, Science,
  324, 1411

\bibitem[{{Arnaud}(1996)}]{arnaud96}
{Arnaud}, K.~A. 1996, in Astronomical Society of the Pacific Conference Series,
  Vol. 101, Astronomical Data Analysis Software and Systems V, ed. G.~H.
  {Jacoby} \& J.~{Barnes}, 17

\bibitem[{Bachetti(2015)}]{bachetti15ascl}
Bachetti, M. 2015, Astrophysics Source Code Library, record ascl:1502.021

\bibitem[{{Bachetti} {et~al.}(2014){Bachetti}, {Harrison}, {Walton},
  {Grefenstette}, {Chakrabarty}, {F{\"u}rst}, {Barret}, {Beloborodov}, {Boggs},
  {Christensen}, {Craig}, {Fabian}, {Hailey}, {Hornschemeier}, {Kaspi},
  {Kulkarni}, {Maccarone}, {Miller}, {Rana}, {Stern}, {Tendulkar}, {Tomsick},
  {Webb}, \& {Zhang}}]{bachetti14}
{Bachetti}, M., {Harrison}, F.~A., {Walton}, D.~J., {et~al.} 2014, \nat, 514,
  202

\bibitem[{{Bahramian} {et~al.}(2014){Bahramian}, {Heinke}, {Sivakoff},
  {Altamirano}, {Wijnands}, {Homan}, {Linares}, {Pooley}, {Degenaar}, \&
  {Gladstone}}]{bahramian14}
{Bahramian}, A., {Heinke}, C.~O., {Sivakoff}, G.~R., {et~al.} 2014, \apj, 780,
  127

\bibitem[{{Baumgardt}(2017)}]{Baumgardt17}
{Baumgardt}, H. 2017, \mnras, 464, 2174

\bibitem[{{Baumgardt} {et~al.}(2002){Baumgardt}, {Hut}, \& {Heggie}}]{BHH02}
{Baumgardt}, H., {Hut}, P., \& {Heggie}, D.~C. 2002, \mnras, 336, 1069

\bibitem[{{Bekki} \& {Freeman}(2003)}]{bekki03}
{Bekki}, K., \& {Freeman}, K.~C. 2003, \mnras, 346, L11

\bibitem[{{Bellini} {et~al.}(2013){Bellini}, {Piotto}, {Milone}, {King},
  {Renzini}, {Cassisi}, {Anderson}, {Bedin}, {Nardiello}, {Pietrinferni}, \&
  {Sarajedini}}]{bellini13}
{Bellini}, A., {Piotto}, G., {Milone}, A.~P., {et~al.} 2013, \apj, 765, 32

\bibitem[{{Bhattacharya} \& {van den Heuvel}(1991)}]{bhattacharya91}
{Bhattacharya}, D., \& {van den Heuvel}, E.~P.~J. 1991, \physrep, 203, 1

\bibitem[{{Bhattacharya} {et~al.}(1992){Bhattacharya}, {Wijers}, {Hartman}, \&
  {Verbunt}}]{bhattacharya92}
{Bhattacharya}, D., {Wijers}, R.~A.~M.~J., {Hartman}, J.~W., \& {Verbunt}, F.
  1992, \aap, 254, 198

\bibitem[{{Bildsten} {et~al.}(1997){Bildsten}, {Chakrabarty}, {Chiu}, {Finger},
  {Koh}, {Nelson}, {Prince}, {Rubin}, {Scott}, {Stollberg}, {Vaughan},
  {Wilson}, \& {Wilson}}]{bildsten97}
{Bildsten}, L., {Chakrabarty}, D., {Chiu}, J., {et~al.} 1997, \apjs, 113, 367

\bibitem[{{Breton} {et~al.}(2007){Breton}, {Roberts}, {Ransom}, {Kaspi},
  {Durant}, {Bergeron}, \& {Faulkner}}]{breton07}
{Breton}, R.~P., {Roberts}, M.~S.~E., {Ransom}, S.~M., {et~al.} 2007, \apj,
  661, 1073

\bibitem[{{Brown} {et~al.}(2016){Brown}, {Cassisi}, {D'Antona}, {Salaris},
  {Milone}, {Dalessandro}, {Piotto}, {Renzini}, {Sweigart}, {Bellini},
  {Ortolani}, {Sarajedini}, {Aparicio}, {Bedin}, {Anderson}, {Pietrinferni}, \&
  {Nardiello}}]{brown16}
{Brown}, T.~M., {Cassisi}, S., {D'Antona}, F., {et~al.} 2016, \apj, 822, 44

\bibitem[{{Buccheri} {et~al.}(1983){Buccheri}, {Bennett}, {Bignami}, {Bloemen},
  {Boriakoff}, {Caraveo}, {Hermsen}, {Kanbach}, {Manchester}, {Masnou},
  {Mayer-Hasselwander}, {Ozel}, {Paul}, {Sacco}, {Scarsi}, \&
  {Strong}}]{buccheri83}
{Buccheri}, R., {Bennett}, K., {Bignami}, G.~F., {et~al.} 1983, \aap, 128, 245

\bibitem[{{Caldwell} {et~al.}(2011){Caldwell}, {Schiavon}, {Morrison}, {Rose},
  \& {Harding}}]{Caldwell+11}
{Caldwell}, N., {Schiavon}, R., {Morrison}, H., {Rose}, J.~A., \& {Harding}, P.
  2011, \aj, 141, 61

\bibitem[{{Cavecchi} {et~al.}(2011){Cavecchi}, {Patruno}, {Haskell}, {Watts},
  {Levin}, {Linares}, {Altamirano}, {Wijnands}, \& {van der Klis}}]{cavecchi11}
{Cavecchi}, Y., {Patruno}, A., {Haskell}, B., {et~al.} 2011, \apjl, 740, L8

\bibitem[{{Chakrabarty} {et~al.}(1997){Chakrabarty}, {Bildsten}, {Grunsfeld},
  {Koh}, {Prince}, {Vaughan}, {Finger}, {Scott}, \& {Wilson}}]{chakrabarty97}
{Chakrabarty}, D., {Bildsten}, L., {Grunsfeld}, J.~M., {et~al.} 1997, \apj,
  474, 414

\bibitem[{{Chilingarian} {et~al.}(2009){Chilingarian}, {Cayatte}, {Revaz},
  {Dodonov}, {Durand}, {Durret}, {Micol}, \& {Slezak}}]{Chilingarian+09}
{Chilingarian}, I., {Cayatte}, V., {Revaz}, Y., {et~al.} 2009, Science, 326,
  1379

\bibitem[{{Chilingarian} \& {Zolotukhin}(2015)}]{CZ15}
{Chilingarian}, I., \& {Zolotukhin}, I. 2015, Science, 348, 418

\bibitem[{{Chilingarian} {et~al.}(2011){Chilingarian}, {Mieske}, {Hilker}, \&
  {Infante}}]{CMHI11}
{Chilingarian}, I.~V., {Mieske}, S., {Hilker}, M., \& {Infante}, L. 2011,
  \mnras, 412, 1627

\bibitem[{{Cohn} {et~al.}(2002){Cohn}, {Lugger}, {Grindlay}, \&
  {Edmonds}}]{cohn02}
{Cohn}, H.~N., {Lugger}, P.~M., {Grindlay}, J.~E., \& {Edmonds}, P.~D. 2002,
  \apj, 571, 818

\bibitem[{{Dabringhausen} {et~al.}(2012){Dabringhausen}, {Kroupa},
  {Pflamm-Altenburg}, \& {Mieske}}]{DKPM12}
{Dabringhausen}, J., {Kroupa}, P., {Pflamm-Altenburg}, J., \& {Mieske}, S.
  2012, \apj, 747, 72

\bibitem[{{Dickey} \& {Lockman}(1990)}]{dl90}
{Dickey}, J.~M., \& {Lockman}, F.~J. 1990, \araa, 28, 215

\bibitem[{{Drinkwater} {et~al.}(2003){Drinkwater}, {Gregg}, {Hilker}, {Bekki},
  {Couch}, {Ferguson}, {Jones}, \& {Phillipps}}]{Drinkwater+03}
{Drinkwater}, M.~J., {Gregg}, M.~D., {Hilker}, M., {et~al.} 2003, \nat, 423,
  519

\bibitem[{{Eggleton}(1983)}]{eggleton83}
{Eggleton}, P.~P. 1983, \apj, 268, 368

\bibitem[{{Esposito} {et~al.}(2016){Esposito}, {Israel}, {Belfiore}, {Novara},
  {Sidoli}, {Rodr{\'{\i}}guez Castillo}, {De Luca}, {Tiengo}, {Haberl},
  {Salvaterra}, {Read}, {Salvetti}, {Sandrelli}, {Marelli}, {Wilms}, \&
  {D'Agostino}}]{esposito16}
{Esposito}, P., {Israel}, G.~L., {Belfiore}, A., {et~al.} 2016, \mnras, 457, L5

\bibitem[{{Evans} {et~al.}(2010){Evans}, {Primini}, {Glotfelty}, {Anderson},
  {Bonaventura}, {Chen}, {Davis}, {Doe}, {Evans}, {Fabbiano}, {Galle}, {Gibbs},
  {Grier}, {Hain}, {Hall}, {Harbo}, {(Helen He}, {Houck}, {Karovska},
  {Kashyap}, {Lauer}, {McCollough}, {McDowell}, {Miller}, {Mitschang},
  {Morgan}, {Mossman}, {Nichols}, {Nowak}, {Plummer}, {Refsdal}, {Rots},
  {Siemiginowska}, {Sundheim}, {Tibbetts}, {Van Stone}, {Winkelman}, \&
  {Zografou}}]{evans10}
{Evans}, I.~N., {Primini}, F.~A., {Glotfelty}, K.~J., {et~al.} 2010, \apjs,
  189, 37

\bibitem[{{Evstigneeva} {et~al.}(2008){Evstigneeva}, {Drinkwater}, {Peng},
  {Hilker}, {De Propris}, {Jones}, {Phillipps}, {Gregg}, \&
  {Karick}}]{evstigneeva08}
{Evstigneeva}, E.~A., {Drinkwater}, M.~J., {Peng}, C.~Y., {et~al.} 2008, \aj,
  136, 461

\bibitem[{{Fabian} {et~al.}(1975){Fabian}, {Pringle}, \& {Rees}}]{fabi75}
{Fabian}, A.~C., {Pringle}, J.~E., \& {Rees}, M.~J. 1975, \mnras, 172, 15

\bibitem[{{Faucher-Gigu{\`e}re} \& {Kaspi}(2006)}]{faucher06}
{Faucher-Gigu{\`e}re}, C.-A., \& {Kaspi}, V.~M. 2006, \apj, 643, 332

\bibitem[{{Ferraro} {et~al.}(2009){Ferraro}, {Dalessandro}, {Mucciarelli},
  {Beccari}, {Rich}, {Origlia}, {Lanzoni}, {Rood}, {Valenti}, {Bellazzini},
  {Ransom}, \& {Cocozza}}]{Ferraro+09}
{Ferraro}, F.~R., {Dalessandro}, E., {Mucciarelli}, A., {et~al.} 2009, \nat,
  462, 483

\bibitem[{{Francis} {et~al.}(2012){Francis}, {Drinkwater}, {Chilingarian},
  {Bolt}, \& {Firth}}]{Francis+12}
{Francis}, K.~J., {Drinkwater}, M.~J., {Chilingarian}, I.~V., {Bolt}, A.~M., \&
  {Firth}, P. 2012, \mnras, 425, 325

\bibitem[{{Galleti} {et~al.}(2004){Galleti}, {Federici}, {Bellazzini}, {Fusi
  Pecci}, \& {Macrina}}]{galleti04}
{Galleti}, S., {Federici}, L., {Bellazzini}, M., {Fusi Pecci}, F., \&
  {Macrina}, S. 2004, \aap, 416, 917

\bibitem[{Ghosh \& Lamb(1978)}]{GhoshLamb}
Ghosh, P., \& Lamb, F.~K. 1978, ApJ, 223, L83

\bibitem[{{Gnedin} {et~al.}(2002){Gnedin}, {Zhao}, {Pringle}, {Fall}, {Livio},
  \& {Meylan}}]{Gnedin+02}
{Gnedin}, O.~Y., {Zhao}, H., {Pringle}, J.~E., {et~al.} 2002, \apjl, 568, L23

\bibitem[{{Harris}(1996)}]{Harris96}
{Harris}, W.~E. 1996, \aj, 112, 1487

\bibitem[{{Heinke} {et~al.}(2006){Heinke}, {Wijnands}, {Cohn}, {Lugger},
  {Grindlay}, {Pooley}, \& {Lewin}}]{heinke06}
{Heinke}, C.~O., {Wijnands}, R., {Cohn}, H.~N., {et~al.} 2006, \apj, 651, 1098

\bibitem[{{H{\'e}non}(1961)}]{heno61}
{H{\'e}non}, M. 1961, Annales d'Astrophysique, 24, 369

\bibitem[{{Heyl} {et~al.}(2015){Heyl}, {Richer}, {Antolini}, {Goldsbury},
  {Kalirai}, {Parada}, \& {Tremblay}}]{Heyl+15}
{Heyl}, J., {Richer}, H.~B., {Antolini}, E., {et~al.} 2015, \apj, 804, 53

\bibitem[{{Hut} {et~al.}(1992){Hut}, {McMillan}, {Goodman}, {Mateo}, {Phinney},
  {Pryor}, {Richer}, {Verbunt}, \& {Weinberg}}]{hut92}
{Hut}, P., {McMillan}, S., {Goodman}, J., {et~al.} 1992, \pasp, 104, 981

\bibitem[{{Ivanova} {et~al.}(2008){Ivanova}, {Heinke}, {Rasio}, {Belczynski},
  \& {Fregeau}}]{ivanova08}
{Ivanova}, N., {Heinke}, C.~O., {Rasio}, F.~A., {Belczynski}, K., \& {Fregeau},
  J.~M. 2008, \mnras, 386, 553

\bibitem[{King \& Lasota(2016)}]{KingLasota16}
King, A., \& Lasota, J.-P. 2016, arXiv, arXiv:1601.03738

\bibitem[{{King}(1966)}]{King66}
{King}, I.~R. 1966, \aj, 71, 64

\bibitem[{{Krist} {et~al.}(2011){Krist}, {Hook}, \& {Stoehr}}]{KHS11}
{Krist}, J.~E., {Hook}, R.~N., \& {Stoehr}, F. 2011, in Society of
  Photo-Optical Instrumentation Engineers (SPIE) Conference Series, Vol. 8127,
  Society of Photo-Optical Instrumentation Engineers (SPIE) Conference Series,
  0

\bibitem[{{Kroupa}(2002)}]{Kroupa02}
{Kroupa}, P. 2002, Science, 295, 82

\bibitem[{{Lanzoni} {et~al.}(2010){Lanzoni}, {Ferraro}, {Dalessandro},
  {Mucciarelli}, {Beccari}, {Miocchi}, {Bellazzini}, {Rich}, {Origlia},
  {Valenti}, {Rood}, \& {Ransom}}]{Lanzoni+10}
{Lanzoni}, B., {Ferraro}, F.~R., {Dalessandro}, E., {et~al.} 2010, \apj, 717,
  653

\bibitem[{Leahy(1987)}]{Leahy87}
Leahy, D.~A. 1987, A{\&}A, 180, 275

\bibitem[{Leahy {et~al.}(1983)Leahy, Darbro, Elsner, Weisskopf, Kahn,
  Sutherland, \& Grindlay}]{Leahy+83}
Leahy, D.~A., Darbro, W., Elsner, R.~F., {et~al.} 1983, ApJ, 266, 160

\bibitem[{{Lyne} {et~al.}(1993){Lyne}, {Biggs}, {Harrison}, \&
  {Bailes}}]{lyne93}
{Lyne}, A.~G., {Biggs}, J.~D., {Harrison}, P.~A., \& {Bailes}, M. 1993, \nat,
  361, 47

\bibitem[{{Lyne} {et~al.}(1996){Lyne}, {Manchester}, \& {D'Amico}}]{lyne96}
{Lyne}, A.~G., {Manchester}, R.~N., \& {D'Amico}, N. 1996, \apjl, 460, L41

\bibitem[{{Manchester} {et~al.}(2005){Manchester}, {Hobbs}, {Teoh}, \&
  {Hobbs}}]{manchester05}
{Manchester}, R.~N., {Hobbs}, G.~B., {Teoh}, A., \& {Hobbs}, M. 2005, \aj, 129,
  1993

\bibitem[{{Michel}(1987)}]{mich87}
{Michel}, F.~C. 1987, \nat, 329, 310

\bibitem[{Mushtukov {et~al.}(2015)Mushtukov, Suleimanov, Tsygankov, \&
  Poutanen}]{Mushtukov+15}
Mushtukov, A.~A., Suleimanov, V.~F., Tsygankov, S.~S., \& Poutanen, J. 2015,
  MNRAS, 447, 1847

\bibitem[{{Narayan} \& {Ostriker}(1990)}]{narayan90}
{Narayan}, R., \& {Ostriker}, J.~P. 1990, \apj, 352, 222

\bibitem[{{Olausen} \& {Kaspi}(2014)}]{olausen14}
{Olausen}, S.~A., \& {Kaspi}, V.~M. 2014, \apjs, 212, 6

\bibitem[{{Osborne} {et~al.}(2001){Osborne}, {Borozdin}, {Trudolyubov},
  {Priedhorsky}, {Soria}, {Shirey}, {Hayter}, {La Palombara}, {Mason},
  {Molendi}, {Paerels}, {Pietsch}, {Read}, {Tiengo}, {Watson}, \&
  {West}}]{osborne01}
{Osborne}, J.~P., {Borozdin}, K.~N., {Trudolyubov}, S.~P., {et~al.} 2001, \aap,
  378, 800

\bibitem[{{Papitto} {et~al.}(2011){Papitto}, {D'A{\`i}}, {Motta}, {Riggio},
  {Burderi}, {di Salvo}, {Belloni}, \& {Iaria}}]{papitto11}
{Papitto}, A., {D'A{\`i}}, A., {Motta}, S., {et~al.} 2011, \aap, 526, L3

\bibitem[{{Patruno} {et~al.}(2012){Patruno}, {Alpar}, {van der Klis}, \& {van
  den Heuvel}}]{patruno12}
{Patruno}, A., {Alpar}, M.~A., {van der Klis}, M., \& {van den Heuvel},
  E.~P.~J. 2012, \apj, 752, 33

\bibitem[{{Patruno} \& {Watts}(2012)}]{patruno12a}
{Patruno}, A., \& {Watts}, A.~L. 2012, ArXiv e-prints, arXiv:1206.2727

\bibitem[{{Peacock} {et~al.}(2010){Peacock}, {Maccarone}, {Knigge}, {Kundu},
  {Waters}, {Zepf}, \& {Zurek}}]{peacock10}
{Peacock}, M.~B., {Maccarone}, T.~J., {Knigge}, C., {et~al.} 2010, \mnras, 402,
  803

\bibitem[{{Peng} {et~al.}(2002){Peng}, {Ho}, {Impey}, \& {Rix}}]{PHIR02}
{Peng}, C.~Y., {Ho}, L.~C., {Impey}, C.~D., \& {Rix}, H.-W. 2002, \aj, 124, 266

\bibitem[{{Perna} {et~al.}(2006){Perna}, {Bozzo}, \& {Stella}}]{perna06}
{Perna}, R., {Bozzo}, E., \& {Stella}, L. 2006, \apj, 639, 363

\bibitem[{{Pfeffer} {et~al.}(2014){Pfeffer}, {Griffen}, {Baumgardt}, \&
  {Hilker}}]{pfeffer14}
{Pfeffer}, J., {Griffen}, B.~F., {Baumgardt}, H., \& {Hilker}, M. 2014, \mnras,
  444, 3670

\bibitem[{{Podsiadlowski} {et~al.}(2002){Podsiadlowski}, {Rappaport}, \&
  {Pfahl}}]{podsiadlowski02}
{Podsiadlowski}, P., {Rappaport}, S., \& {Pfahl}, E.~D. 2002, \apj, 565, 1107

\bibitem[{{Pooley} {et~al.}(2003){Pooley}, {Lewin}, {Anderson}, {Baumgardt},
  {Filippenko}, {Gaensler}, {Homer}, {Hut}, {Kaspi}, {Makino}, {Margon},
  {McMillan}, {Portegies Zwart}, {van der Klis}, \& {Verbunt}}]{pooley03}
{Pooley}, D., {Lewin}, W.~H.~G., {Anderson}, S.~F., {et~al.} 2003, \apjl, 591,
  L131

\bibitem[{{Prager} {et~al.}(2016){Prager}, {Ransom}, {Freire}, {Hessels},
  {Stairs}, {Arras}, \& {Cadelano}}]{prager16}
{Prager}, B., {Ransom}, S., {Freire}, P., {et~al.} 2016, ArXiv e-prints,
  arXiv:1612.04395

\bibitem[{{Radhakrishnan} \& {Srinivasan}(1982)}]{radh82}
{Radhakrishnan}, V., \& {Srinivasan}, G. 1982, Current Science, 51, 1096

\bibitem[{Ransom(2001)}]{RansomThesis}
Ransom, S.~M. 2001, PhD thesis, Harvard University

\bibitem[{{Richstone} \& {Tremaine}(1986)}]{RT86}
{Richstone}, D.~O., \& {Tremaine}, S. 1986, \aj, 92, 72

\bibitem[{{Riess} {et~al.}(2012){Riess}, {Fliri}, \& {Valls-Gabaud}}]{riess12}
{Riess}, A.~G., {Fliri}, J., \& {Valls-Gabaud}, D. 2012, \apj, 745, 156

\bibitem[{{Rosen} {et~al.}(2015){Rosen}, {Webb}, {Watson}, {Ballet}, {Barret},
  {Braito}, {Carrera}, {Ceballos}, {Coriat}, {Della Ceca}, {Denkinson},
  {Esquej}, {Farrell}, {Freyberg}, {Gris{\'e}}, {Guillout}, {Heil},
  {Law-Green}, {Lamer}, {Lin}, {Martino}, {Michel}, {Motch}, {Nebot
  Gomez-Moran}, {Page}, {Page}, {Page}, {Pakull}, {Pye}, {Read}, {Rodriguez},
  {Sakano}, {Saxton}, {Schwope}, {Scott}, {Sturm}, {Traulsen}, {Yershov}, \&
  {Zolotukhin}}]{rosen15}
{Rosen}, S.~R., {Webb}, N.~A., {Watson}, M.~G., {et~al.} 2015, ArXiv e-prints,
  arXiv:1504.07051

\bibitem[{{Seth} {et~al.}(2014){Seth}, {van den Bosch}, {Mieske}, {Baumgardt},
  {Brok}, {Strader}, {Neumayer}, {Chilingarian}, {Hilker}, {McDermid},
  {Spitler}, {Brodie}, {Frank}, \& {Walsh}}]{Seth+14}
{Seth}, A.~C., {van den Bosch}, R., {Mieske}, S., {et~al.} 2014, \nat, 513, 398

\bibitem[{{Shaw Greening} {et~al.}(2009){Shaw Greening}, {Barnard}, {Kolb},
  {Tonkin}, \& {Osborne}}]{greening09}
{Shaw Greening}, L., {Barnard}, R., {Kolb}, U., {Tonkin}, C., \& {Osborne},
  J.~P. 2009, \aap, 495, 733

\bibitem[{{Siegel} {et~al.}(2007){Siegel}, {Dotter}, {Majewski}, {Sarajedini},
  {Chaboyer}, {Nidever}, {Anderson}, {Mar{\'{\i}}n-Franch}, {Rosenberg},
  {Bedin}, {Aparicio}, {King}, {Piotto}, \& {Reid}}]{siegel07}
{Siegel}, M.~H., {Dotter}, A., {Majewski}, S.~R., {et~al.} 2007, \apjl, 667,
  L57

\bibitem[{{Spitzer}(1987)}]{spitzer87}
{Spitzer}, L. 1987, {Dynamical evolution of globular clusters}

\bibitem[{{Spitzer} \& {Hart}(1971)}]{spitzer71}
{Spitzer}, Jr., L., \& {Hart}, M.~H. 1971, \apj, 164, 399

\bibitem[{{Strader} {et~al.}(2011){Strader}, {Caldwell}, \& {Seth}}]{SCS11}
{Strader}, J., {Caldwell}, N., \& {Seth}, A.~C. 2011, \aj, 142, 8

\bibitem[{{Supper} {et~al.}(2001){Supper}, {Hasinger}, {Lewin}, {Magnier}, {van
  Paradijs}, {Pietsch}, {Read}, \& {Tr{\"u}mper}}]{supper01}
{Supper}, R., {Hasinger}, G., {Lewin}, W.~H.~G., {et~al.} 2001, \aap, 373, 63

\bibitem[{Taylor(1992)}]{Taylor92}
Taylor, J.~H. 1992, Philosophical Transactions: Physical Sciences and
  Engineering, 341, 117

\bibitem[{{Trudolyubov} {et~al.}(2005){Trudolyubov}, {Kotov}, {Priedhorsky},
  {Cordova}, \& {Mason}}]{trudolyubov05}
{Trudolyubov}, S., {Kotov}, O., {Priedhorsky}, W., {Cordova}, F., \& {Mason},
  K. 2005, \apj, 634, 314

\bibitem[{{Trudolyubov} \& {Priedhorsky}(2004)}]{trudolyubov04}
{Trudolyubov}, S., \& {Priedhorsky}, W. 2004, \apj, 616, 821

\bibitem[{{Trudolyubov} \& {Priedhorsky}(2008)}]{trudolyubov08}
{Trudolyubov}, S.~P., \& {Priedhorsky}, W.~C. 2008, \apj, 676, 1218

\bibitem[{{van der Klis}(1998)}]{vand98}
{van der Klis}, M. 1998, Advances in Space Research, 22, 925

\bibitem[{{Verbunt}(2003)}]{verbunt03}
{Verbunt}, F. 2003, in Astronomical Society of the Pacific Conference Series,
  Vol. 296, New Horizons in Globular Cluster Astronomy, ed. G.~{Piotto},
  G.~{Meylan}, S.~G. {Djorgovski}, \& M.~{Riello}, 245

\bibitem[{{Verbunt} \& {Freire}(2014)}]{verbunt14}
{Verbunt}, F., \& {Freire}, P.~C.~C. 2014, \aap, 561, A11

\bibitem[{{Verbunt} \& {Hut}(1987)}]{verbunt87}
{Verbunt}, F., \& {Hut}, P. 1987, in IAU Symposium, Vol. 125, The Origin and
  Evolution of Neutron Stars, ed. D.~J. {Helfand} \& J.-H. {Huang}, 187

\bibitem[{{Verbunt} {et~al.}(1990){Verbunt}, {Wijers}, \& {Burm}}]{verbunt90}
{Verbunt}, F., {Wijers}, R.~A.~M.~J., \& {Burm}, H.~M.~G. 1990, \aap, 234, 195

\bibitem[{{White} \& {Angelini}(2001)}]{white01}
{White}, N.~E., \& {Angelini}, L. 2001, \apjl, 561, L101

\bibitem[{{Wijnands} \& {van der Klis}(1998)}]{wijn98}
{Wijnands}, R., \& {van der Klis}, M. 1998, \nat, 394, 344

\bibitem[{{Yungelson} {et~al.}(2002){Yungelson}, {Nelemans}, \& {van den
  Heuvel}}]{yungelson02}
{Yungelson}, L.~R., {Nelemans}, G., \& {van den Heuvel}, E.~P.~J. 2002, \aap,
  388, 546

\end{thebibliography}
\bibliographystyle{apj}

\appendix

\section{Improved period search and error analysis for the epoch folding technique}

\subsection{Error analysis for the epoch folding of a single observations}
\label{sec:epoch_folding}

Here we follow \citet{Leahy+83,Leahy87} in order to
perform the period search using the epoch folding technique and make an
additional step and estimate the period determination uncertainty from
statistical considerations.

Let us consider a harmonic signal with a period $P$ over some constant 
background so that the pulse shape is expressed as:
\begin{align}
f(t) = a + b \sin t \nonumber \\
a = N_{\gamma}/T
\label{eqpulse}
\end{align}
\noindent where $b$ is the pulse amplitude and $a$ is the background value
that we estimate from the total number of photons $N_{\gamma}$ registered
during the total exposure time $T$.  Here we assume that the exposure
filling factor is 100~per~cent, i.e.  no gaps took place during the exposure
time due to e.g.  soft proton flares.

If we now perform the epoch folding with a slightly different period
$P+\Delta P, \Delta P \ll P$, it will cause the phase shift of the last pulse:
\begin{equation}
\Delta \phi = \frac{2 \pi T \Delta P}{P^2}
\end{equation}

Hereafter, we take the continuum limit and replace all sums with integrals.
The discretization does not change the final results much because the 
reduction in the method sensitivity is only 3.3~per~cent for $n=10$ bins per
phase and 0.8~per~cent for $n=20$ used by us here \citep{Leahy87}.
Then we can estimate the phase smeared folded pulse shape as a function 
of $\Delta \phi$:
\begin{align}
f_1(t,\Delta \phi) = \frac{1}{\Delta \phi}\int_0^{\Delta \phi} (a + b \sin
(t-\tau)) d\tau = \nonumber \\
a + \frac{2 b}{\Delta \phi} \sin \frac{\Delta \phi}{2} \sin (t-\frac{\Delta
\phi}{2})
\label{eqpulseshift}
\end{align}

Now we can compute the $S$ statistics \citep{Leahy+83} as
\begin{align}
S(\Delta \phi) = \int_0^{2\pi} \frac{(f_1(t, \Delta \phi) - a)^2}{f_1(t,
\Delta \phi)} dt \approx 
\frac{1}{a} \int_0^{2\pi} (f_1(t, \Delta \phi) - a)^2 dt = \nonumber \\
\frac{2\pi b^2}{a \Delta \phi^2} (1 - \cos \Delta \phi) = \frac{4\pi b^2}{a
\Delta \phi^2} \sin^2 \frac{\Delta \phi}{2} 
\label{eqsphi}
\end{align}
\noindent For the simplicity of the computation, here we assume that the 
pulse is shallow (e.g. $b \ll a$) and hence we take $1/a$ outside the
integral. More general analytical calculation is bulky and does not 
change the result very much because it depends weakly on the $b/a$ ratio 
as $\sim \sqrt{1-(b/a)^2}$.

Now keeping in mind that $S$ is in fact the $\chi^2$ statistics, we can
estimate the period uncertainty by solving the equation $S(\Delta \phi) =
S(0) - 1$. The Taylor expansion of Eq.~\ref{eqsphi} to the 4-th power 
on $\Delta \phi$ yields:
\begin{equation}
\frac {\pi b^2}{a} (1-\frac{\Delta \phi^2}{12}) = \frac {\pi b^2}{a} - 1
\end{equation}
\noindent Solving it for $\Delta \phi$ and introducing the pulse depth 
$A = b/a$ \citep[as in][]{Leahy87}:
\begin{equation}
\Delta P = \sqrt{\frac{3}{\pi^3}}\frac{P^2}{A \sqrt{N_{\gamma} T}}
\label{eqdeltap}
\end{equation}

\subsection{Analytic formulation for the epoch folding of coherent pulsations
in two observations}

Now let us consider a harmonic signal with a period $P$ observed in two
observations with exposure times $T$ and $T/n$ (without the loss of
generality we take a real number $n \ge 1$). The second observation starts
at the moment $mT$ ($m$ is a real number, $m \ge 1$). Then, the folded phase
shape resulting from the sum of the two observations becomes:
\begin{align}
f_1(t,\Delta \phi,m,n) = \frac{1}{\Delta \phi} (\int_0^{\Delta
\phi} (a + b \sin
(t-\tau)) d\tau + \int_{m \Delta \phi}^{(m+1/n) \Delta \phi} (a + b \sin (t-\tau))
d\tau)
\label{eqpulseshift2}
\end{align}
\noindent Here the second integral has a multiplier $(1/\Delta \phi)$ rather
than $(n/\Delta \phi)$ because it will contribute as $(1/n)$ to the total
pulse. Omitting bulky computations and trigonometric transformations, the 
$S$ statistics computed in the shallow pulse approximation ($b\ll a$, see above)  
becomes:
\begin{align}
S(\Delta \phi,m,n) = \frac{4\pi b^2}{a
\Delta \phi^2} (\sin^2 \frac{\Delta \phi}{2} + \sin^2 \frac{\Delta \phi}{2n}
+ 2 \sin \frac{\Delta \phi}{2} \sin \frac{\Delta \phi}{2n} \cos(\frac{\Delta \phi}{2}
- \frac{\Delta \phi}{2n} - m \Delta \phi))
\label{eqsphi2}
\end{align}

\begin{figure}
\includegraphics[width=\hsize]{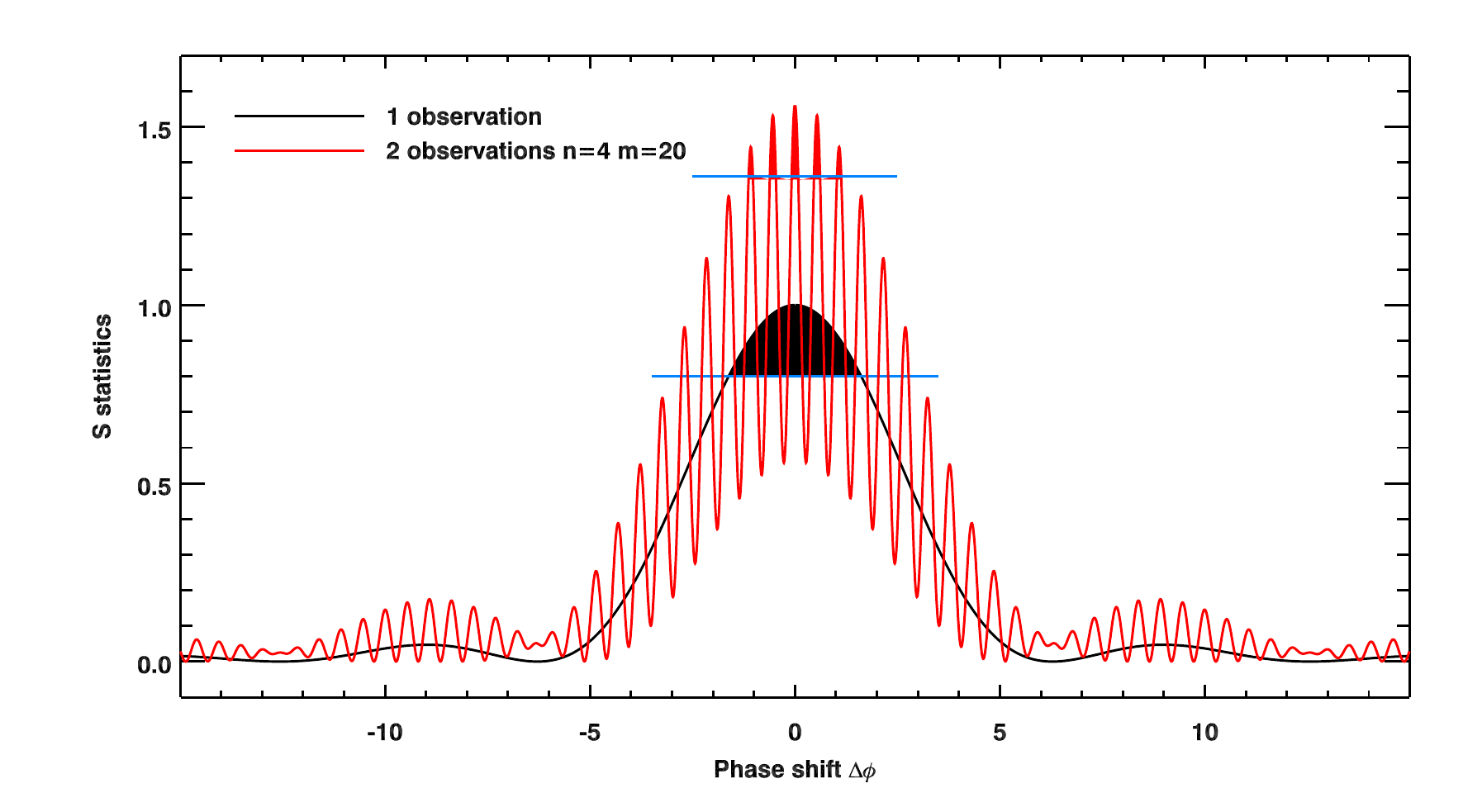}
\caption{Examples of the $S$ statistics for the period search (without the
stochastic component) for one observation (black line) with a total exposure
$T$ shown as black solid line and for two observations (red line) with
exposure times $T_1=T$ and $T_2=T/4$ with the second observations starting
at the moment $T_s=12 T$. Shaded areas show examples of confidence regions:
for the red curve it contains two secondary peaks on either side of the
primary one.
\label{fig:splot_app}}
\end{figure}

This expression is non-negative for any $\Delta \phi$ and it has several
properties of interest for our analysis.  It is virtually identical to the
light pattern formed by the double slit diffraction of a coherent source on
the slits of unequal widths.  Adding the second dataset separated from the
first one in time introduces the modulation of the $S$ statistics from a
single observation (Eq.~\ref{eqsphi}) with the amplitude $(1+1/n)^2$ and the
high frequency $m/2\pi$ (see Fig.~\ref{fig:splot_app}).  Then, depending on the
statistics defined by the total number of registered photons, the 1$\sigma$
confidence region for the period $P$ may either shrink into a single
modulated peak so that the period uncertainty $\Delta P$ will be given by
the equation similar to Eq.~\ref{eqdeltap}:
\begin{equation}
\Delta P_{\mathrm{peak}} = \sqrt{\frac{3}{\pi^3}}\frac{P^2 \sqrt{1+1/n}}{A \sqrt{N_{\gamma} T}
(m+1/n)}
\label{eqdeltap2high}
\end{equation}
\noindent or cover several secondary peaks in which case a single value of
$\Delta P$ becomes meaningless because the real $P$ value may reside in one
of several secondary peaks in the confidence region.  The multiple solution
situation occurs when the phase shift $\Delta \phi$ that corresponds to the
period uncertainty $\Delta P$ from Eq.~\ref{eqsphi} (with $T$ corrected to
the total exposure time $T_{\mathrm{tot}}=T(1+1/n)$) exceeds the phase
distance between secondary peaks: $\Delta \phi > 2\pi/m$ assuming $m \ll 1$.  
Hence, we can estimate the number of secondary peaks $n_{\mathrm{sec}}$ within 
the upper envelope of the $S$ distribution by equating the phase shift to 
$2\pi n_{\mathrm{sec}}/m$:
\begin{equation}
n_{\mathrm{sec}} = 2 \lfloor \sqrt{\frac{3}{\pi^3}}\frac{m}{A \sqrt{a}} \rfloor
\end{equation}
\noindent This quantity depends on the count rate $a =
N_{\gamma}/T_{\mathrm{tot}}$ and not on the actual counts. This conclusion
looks counter-intuitive at first, however, it is trivially explained by the fact that
we work in phase space. Therefore, when the pulse statistics grows 
because of the increased exposure time, the main peak becomes narrower in 
terms of $\Delta P$ and if we keep the separation between observations
constant, $m$ will decrease. In the case of XB091D $a\approx0.1$ and 
$A\approx0.3$, we will get $\sim3m$ secondary peaks on either side of 
the primary one if we combine two observations. 
Another important property of Eq.~\ref{eqsphi2} is that one can use the
measured modulation amplitude of the $S$ statistics for the period search in
two observations in order to estimate the possible period variation.  If the
observed modulation stays close to the value predicted by Eq.~\ref{eqsphi2}
plus the stochastic component $S_{noise}$ that a $\chi^2_{n-1}$ statistics
with mean $(n-1)$ \citep{Leahy87}, we can conclude that the period change
between two observations is undetectable.

All the computations presented above can be trivially generalized to the
case of $N_{\mathrm{obs}}>2$.  However, all expressions become very lengthy
and difficult to understand.  The envelope size for the confidence region
that might still contain multiple narrow peaks (1$\sigma$) can be estimated
as:
\begin{equation}
\Delta P_{\mathrm{mult}} = \sqrt{\frac{3}{\pi^3}}\frac{P^2}{A
\sqrt{N_{\gamma \mathrm{tot}}
T_{\mathrm{tot}}}}
\label{eqdeltapmult}
\end{equation}
\noindent where $T_{\mathrm{tot}}$ indicates the total good time interval
(GTI) and $N_{\gamma \mathrm{tot}}$ is the total number of detected photons.

\subsection{Regularization of the pulse shape}

We can improve the period search procedure by assuming the pulse shape to be
smooth. This idea is similar to regularization techniques used in the
image/signal reconstruction and therefore we can exploit similar
mathematical methods. As long as the period is determined by searching
the maximum of the $S$ statistics, we introduce the penalization factor that
depends on the pulse shape $p(\phi)$ as:
\begin{equation}
S_{\mathrm{reg}} = S (1+k \int_0^{2\pi} (\frac{d^2p(\phi)}{d\phi^2})^2 
d\phi)^{1/2}
\end{equation}
\noindent Here $k$ is a positive coefficient that defines the degree of
regularization. Numerically, the second derivative is computed for a discrete
pulse shape as $\frac{3}{2}(box(p,3) - p)$ where $box(p,3)$ denotes the boxcar
smoothing of the pulse $p$ with a window of 3 pixels. 

\begin{figure}
\includegraphics[width=\hsize]{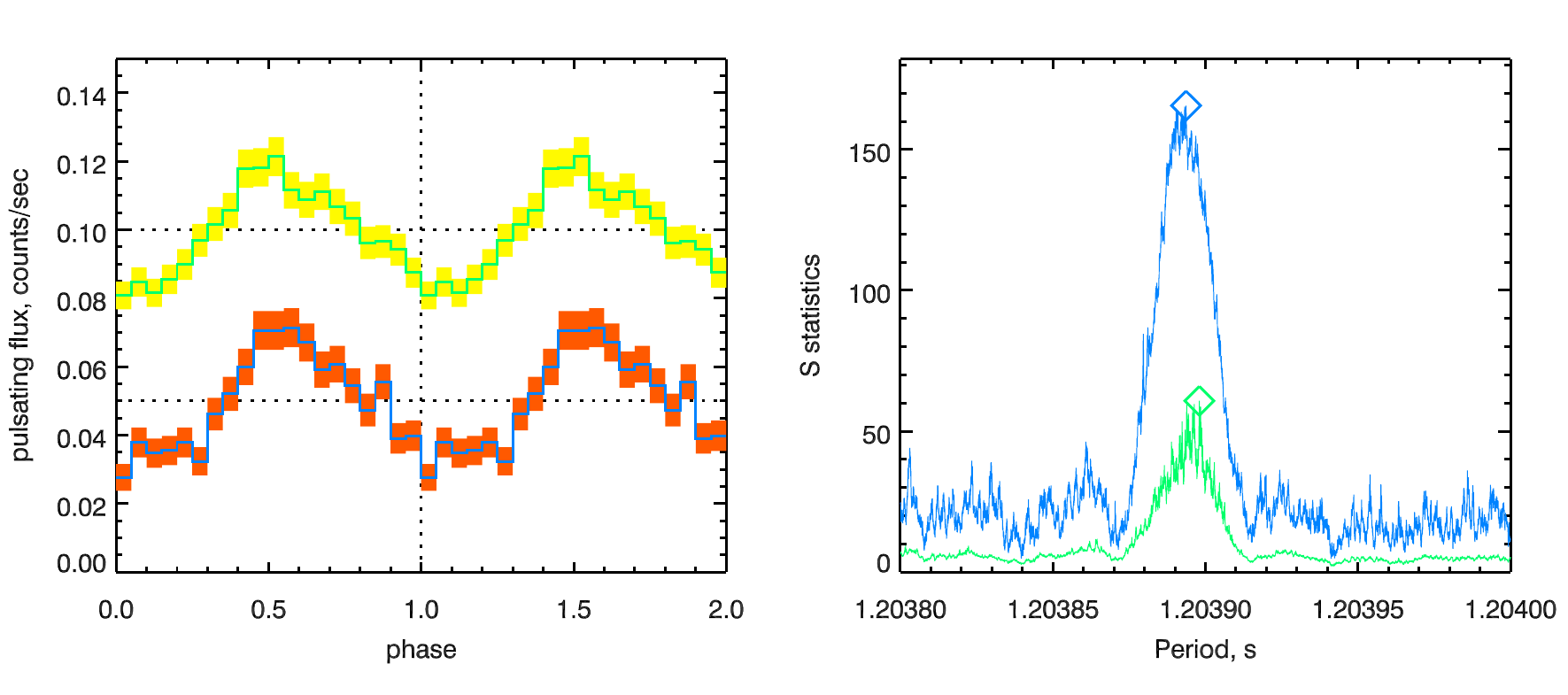}
\caption{The pulse profile recovered from the observation collected on
2002-01-06 (ObsID: 0112570101). ({\it Left}) The yellow and red lines are pulse shapes obtained with and without regularization
respectively.
({\it Right}) The green and blue lines represent the S statistics versus spin period for the search with and without regularization respectively.
\label{fig:pulse_prof_2002}}
\end{figure}

Applying the regularization to real data results in much smoother pulse
shapes while the periods always stay within the uncertainties predicted from
the $S$ statistics or the analytic formulations provided above
(Eq.~\ref{eqdeltapmult}).  We stress that we do not smooth the actual pulse
shape but rather bias the period search statistically towards smoother
shapes.  In Fig.~\ref{fig:pulse_prof_2002} (left panel) we show an example
of the recovered pulse shape from the dataset obtained in January of 2002 (ObsID:
0112570101) with and without regularization using $k=(2/3)^2$. The right
panel shows the $S$ statistics without (blue) and with (green)
regularization obtained from the folded epoch period search in 20 phase
bins. Therefore, the $S$ statistics includes a stochastic contribution with
the variance $<S_{\mathrm{rand}}> = 19$. It is clear from the plot that the
regularized solution lies within the $1 \sigma$ confidence area, i.e. the
period values are consistent within statistical uncertainties, however, the
recovered pulse shape looks substantially smoother in the regularized case.

\end{document}